\documentclass[11pt]{article}
\usepackage{graphicx}
\usepackage{amssymb}
\usepackage{amsmath}
\usepackage{amsthm}
\usepackage{amscd}
\usepackage{hyperref}
\usepackage{esint}
\usepackage{color}
\usepackage[ruled]{algorithm2e}
\usepackage{array}

\SetKwRepeat{Do}{do}{while}

\allowdisplaybreaks
    
\newcommand{\bitradius}{r} 
\newcommand{\bitlength}{b} 
\newcommand{\headradiusA}{h} 
 
\newcommand{\millVolume}{V}
\newcommand{\partDomain}{\Omega}
 
\newcommand{\surfacePt}{x} 
\newcommand{\sdf}{\phi}
\newcommand{\sdfGrad}{\nabla\sdf}
\newcommand{\TempField}{T}
\newcommand{\TempFieldGrad}{\nabla\TempField}
\newcommand{\intPt}{x_{\mathit{hit}}}
\newcommand{\speed}{v}
\newcommand{\maxSpeed}{v_{max}}
\newcommand{\posVal}{\alpha}
\newcommand{\velocity}{\mathbf{\speed}} 
 
\newcommand{\advectTime}{t}
\newcommand{\filter}{\eta} 
\newcommand{\normal}{\mathbf{n}}
\newcommand{\millingDir}{\mathbf{m}}
\newcommand{\millingDirSet}{\mathcal {M}}

\newcommand{\toolTip}{\mathbf{p}}
\newcommand{\heatPt}{\mathbf{q}} 
\newcommand{\heatEps}{\varepsilon} 

\newcommand{\offset}{o}
\newcommand{\hitPt}{x_{\mathit{hit}}}

\newcommand{\R}{\mathbb{R}}

\newcommand{\eps}{\varepsilon}

\newtheorem{thm}{Theorem}

\theoremstyle{definition}
\newtheorem{defn}[thm]{Definition}

\newtheoremstyle{rmk}{5pt}{5pt}{}{}{\scshape}{:}{.5em}{}
\theoremstyle{rmk}

\newcommand{\mylabel}
	{\label}

\newcounter{fix}
\begin{document}

\renewcommand{\baselinestretch}{1.125}
\normalsize

\centerline{\LARGE A Subtractive Manufacturing Constraint} 
\medskip
\centerline{\LARGE for Level Set Topology Optimization}
\bigskip
\centerline{Nigel Morris, Adrian Butscher, Francesco Iorio}
\bigskip
\centerline{3 July 2019}
\bigskip

We present a method for enforcing manufacturability constraints in generated parts such that they will be automatically ready for fabrication using a subtractive approach. We primarily target multi-axis CNC milling approaches but the method should generalize to other subtractive methods as well. To this end we take as user input: the radius of curvature of the tool bit, a coarse model of the tool head and optionally a set of milling directions. This allows us to enforce the following manufacturability conditions: 1) surface smoothness such that the radius of curvature of the part does not exceed the milling bit radius, 2) orientation such that every part of the surface to be milled is visible from at least one milling direction, 3) accessibility such that every surface patch can be reached by the tool bit without interference with the tool or head mount. We will show how to efficiently enforce the constraint during level set based topology optimization modifying the advection velocity such that at each iteration the topology optimization maintains a descent optimization direction and does not violate any of the manufacturability conditions. This approach models the actual subtractive process by carving away material accessible to the machine at each iteration until a local optimum is achieved. 

\section{Introduction}

The aim of this work is to optimally synthesize the geometry of a mechanical part under a specified set of loading conditions and constraints such that the part can be successfully and accurately manufactured using a subtractive process. This process typically begins with a solid block of stock material and gradually removes material from the block until the remaining material has the shape of the designed part. One of the most common subtractive approaches is called ``milling'' and uses rotary cutters otherwise known as ``tools,'' ``end-mills'' or ``bits'' to remove the material \cite{Brown14}. This process inherently limits the types of shapes that can be manufactured, since the milling machine must be able to hold the part rigidly and the rotary bit must be able to access the material surface without interference. Other important considerations must also be taken into account such as vibration of the part during material removal and stresses on the bit itself by the milling process. We focus the scope of this work on the problem of synthesizing optimal shapes whose surface is entirely accessible by a user-specified milling machine and tool.

The process of synthesizing optimal shapes given specified boundary conditions including loads and restraints is generally called \emph{structural topology optimization} and there are two primary categories, density-based and boundary-based \cite{Sigm13}. Density-based approaches discretize the volume of the part and assign a density to each discrete cell, then the densities are driven toward solid (1) and empty (0) while supporting the specified boundary conditions \cite{Bend88}. Boundary-based approaches instead track the shape of the external interface of the solid part and move this boundary incrementally towards optimality. In order to enforce the manufacturability constraint, having a precise knowledge of the boundary is advantageous so we make use of the latter approach. In addition, the method we employ uses an implicit representation known as a level set \cite{Wang03,Alla04} to track the boundary. 

\subsection{Related Work}

In \cite{Ma15}, the problem of manufacturability is tackled by a different approach. During topology optimization, a set of millable extrusions are fit to the geometry generated by the level set method and this approximate geometry is readily manufacturable. However the final optimized geometry without constraints in general cannot be represented by a set of extrusions and thus some level of optimality is lost in this approximation process. 

In \cite{Vatan16,Wei16,Zhu12}, topology optimization using a density-based method and a so-called `projection' or `visibility map' is used to constrain output to manufacturable parts. Manufacturable here is simply defined as every element in the density volume must have no occlusion in the milling direction. Occlusion means that there must be no density higher than the element's density in elements along the specified milling direction. Depending on the manufacturability technique, elements violating the constraint are either filled or removed. While this approach is convenient, unfortunately density-based methods do not converge to solid and empty elements until the later stages of the algorithm and so the constraint will not be physically accurate until this point of the optimization. In \cite{Vatan16,Wei16} there is no explicit modeling of the subtractive bit or holder for milling manufacturability. In \cite{Zhu12} the subtractive bit is modeled as a cylinder with hemi-spherical cap, but there is no constraint on the bit length or model of the holder. In \cite{Lang19} the `projection' method was extended to include multi-axis milling constraints by using affine transformations to rotate the densities to align with the milling directions before performing an aggregated `projection.' They also include a method for modeling the bit and holder, and model the effect of these on the optimized geometry. In this method the user must specify the set of possible milling directions from which material can be removed.

Our method works in a similar way to \cite{Xia10,Wang17} where a level set-based topology optimization is performed and the level set velocity function is constrained in order to enforce the manufacturing property. In contrast to \cite{Xia10,Wang17} which constrain the part to be manufactured with a two-sided cast, our manufacturing method is a more general subtractive constraint with a user defined tool from either a fixed set of directions or automatically determined 5-axis subtraction.

\subsection{Contributions}

\paragraph{Mill geometry.} To our knowledge this is the first level set-based method to incorporate accessibility of a user defined physical tool bit and mill head.

\vspace{-1ex}
\paragraph{Multi-axis constraint.} We find the most accessible milling direction from a user supplied set or automatically choose the best milling direction to simulate 5-axis milling.

\vspace{-1ex}

\paragraph{Compatible with a shape gradient-based formulation.} We impose our milling constraint by modifying the shape gradients used in the level set-based topology optimization algorithm in such a way that a descent direction is maintained in every optimization iteration.

\section{Definition of the Milling Constraint}
\label{sec:constraint}

\begin{figure}[h]
	\begin{center}
		\small
		\includegraphics[width=\textwidth]{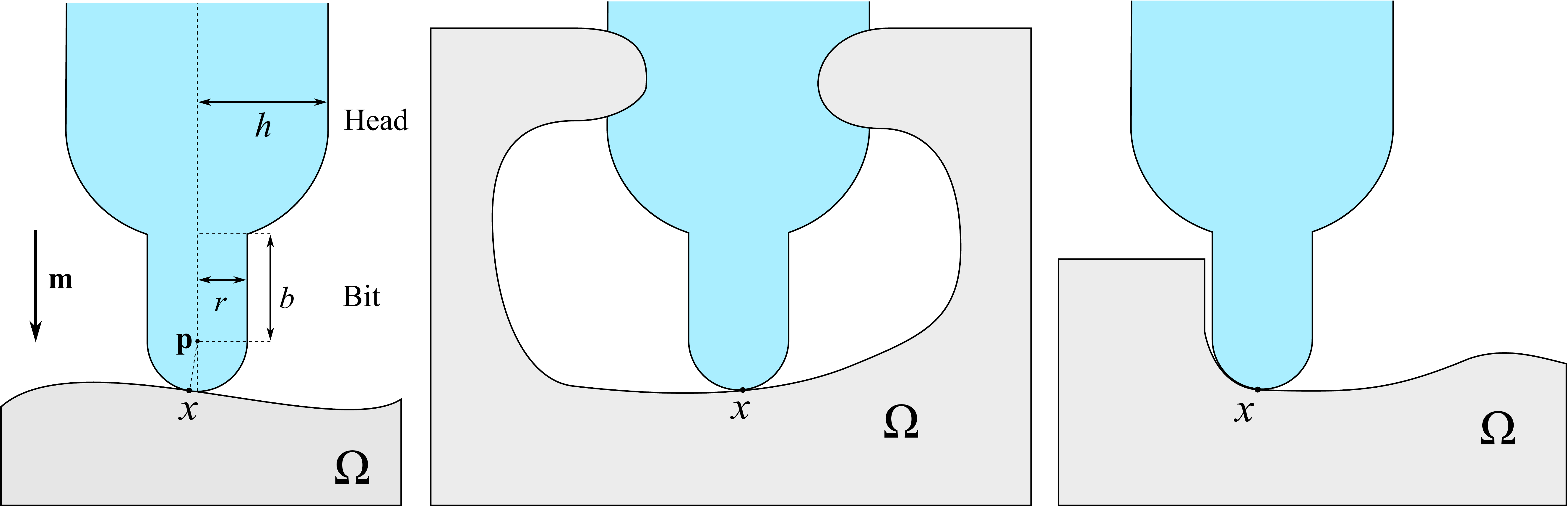}  
		\caption{\textbf{Left:} Milling geometry definitions.  \textbf{Middle:} Example of inaccessible partial cavity. \textbf{Right:} Example of a millable surface point. }
		\label{fig:constraintEx}
	\end{center}
\end{figure}

We consider a tool bit (or end mill) and head that is defined using the following parameters: bit radius $\bitradius$ and bit length $\bitlength$ define the shape of the tool bit represented by a cylinder capped with a hemisphere oriented in the milling direction $\millingDir$, and the head radius $\headradiusA$ defines another such capped cylinder for the tool holder and head.  We assume that the end of the head cylinder extends infinitely far away from the surface.  Additionally, for 3-axis milling, we allow the user to specify a set of \emph{milling directions} $\millingDirSet$ from which the mill can approach the surface of the part. Whereas for 5-axis milling, we consider that the milling tool can be oriented to approach the surface from an arbitrary direction and our algorithm automatically chooses the milling direction.  See Figure \ref{fig:constraintEx} for an illustration of the quantities defined above.

For the purpose of this work, a part is millable if all its surface points are accessible in the sense that a tool bit as described above can be brought in from infinity along an allowed milling direction until it touches the surface point, making no intersection with the interior of the part.  

\begin{defn}
\label{def:millability}
Let $\Omega$ be a shape with boundary surface $\partial \Omega$. A point $\surfacePt \in \partial \Omega$ is \emph{millable} if there exists at least one milling direction for which the tool bit touches $\surfacePt$ and the tool bit and head as defined in the above text do not intersect with the interior of $\partDomain$.  The boundary surface as a whole is millable if all its points are millable.
\end{defn}

\noindent See Figure \ref{fig:constraintEx} for examples of millable and non-millable points.

\section{Topology Optimization with Millability Constraints}
\label{sec:algorithm}

\subsection{Topology Optimization}
\label{sec:to}

Topology optimization problems can be formulated abstractly as the search for an optimal shape belonging to a class of \emph{admissible shapes}, denoted herein by $\mathit{Adm}$.  The specific notion of admissibility that we will use in this paper will be defined below (see Definition \ref{def:admissibility}).  The sought-after shape $\Omega$ solves the constrained optimization problem
\begin{equation}
	\label{eqn:toproblem}
	\begin{aligned}
		\min_{\Omega \in \mathit{Adm}}  \qquad &\mathcal F(\Omega) \\
		\mbox{s.t. } \qquad &\mathcal G_i(\Omega) \leq 0 \quad \forall \; i=1, \ldots, k
	\end{aligned}
\end{equation}
in a suitable sense --- namely, we content ourselves with a \emph{feasible} (i.e.\ constraint-satisfying) \emph{local minimum} of \eqref{eqn:toproblem}.  Here $\mathcal F, \mathcal G_1, \ldots, \mathcal G_k : \mathit{Adm} \rightarrow \R$ are the objective and one or more inequality constraint functions.  In structural topology optimization problems, at least one of these functions is formulated in terms of the solutions of  linear elasto-static partial differential equations with respect to one or more load cases.  Each load case consists of a surface traction (a.k.a.\ Neumann boundary condition) applied to a subset of $\partial \Omega$ and a prescribed displacement (a.k.a.\ Dirichlet boundary condition) applied to another subset of $\partial \Omega$.   

We will use level set-based topology optimization \cite{Alla04,Wang03} with the \emph{augmented Lagrangian algorithm} \cite{nocedal} for solving the problem \eqref{eqn:toproblem}.  This is an iterative strategy, whereby shapes are represented as the interior regions delimited by the zero-contour of piecewise smooth \emph{level set functions} defined on an ambient design domain.  The level set function for the shape at each iteration is updated in a manner that improves the optimization objective and reduces the constraint violation until a feasible, locally optimal admissible shape is achieved.  Specifically, the update is performed by integrating a well-chosen \emph{boundary normal speed function} for a small pseudo-time in each iteration by means of the standard level set Hamilton-Jacobi equation.  (This update procedure is known as \emph{advection} with respect to the chosen boundary normal speed.)  In our case, the speed function on the boundary of the shape in the current iteration equals the negative \emph{shape gradient}  of the \emph{augmented Lagrangian} which is a well-chosen algebraic combination of the objective and constraint functions. The speed function is then extended to a narrow band of the boundary by insisting that the extension be constant along the normal direction (to achieve this we solve the normal extension Hamilton-Jacobi equation for a time proportional to the width of the band).  If $\mathcal L$ denotes the augmented Lagrangian and $d\mathcal L$ denotes its shape gradient at $\Omega$, then we now have a speed function $\speed : \R^3 \rightarrow \R$ such that
\begin{equation}
	\label{eqn:speedchoice}
	\speed(\surfacePt) = - d\mathcal L(\surfacePt)
\end{equation}
for all $\surfacePt$ belonging to the boundary $\partial \Omega$ of the geometry $\Omega$.   Additionally, we set $v$ to zero on any part of the boundary of $\Omega$ we wish to hold fixed, such as the surface patches where the non-zero surface loads and prescribed displacements of each load case are applied.  The chosen speed function is a \emph{descent direction} for the augmented Lagrangian since it can be shown \cite{Zole11} that
\begin{equation}
	\label{eqn:descent}
	\mathcal L(\Omega_\eps) = \mathcal L(\Omega) + \eps \int_{\partial \Omega} v\, d\mathcal L + \mathcal O(\eps^2)  = \mathcal L(\Omega) - \eps \int_{\Gamma} \big( d\mathcal L  \big)^2 + \mathcal O(\eps^2) \, ,
\end{equation}
where $\Omega_\eps$ is the advected geometry at the pseudo-time $\eps$ and $\Gamma$ is the subset of $\partial \Omega$ that is allowed to move.  This formula implies that $\mathcal L(\Omega_\eps) < \mathcal L(\Omega)$ for sufficiently small $\eps$.  After iteratively applying this procedure and also updating the penalty coefficients and the Lagrange multipliers used to define the augmented Lagrangian, the shape converges to a local solution of \eqref{eqn:toproblem}.

\subsection{Admissible Shapes}

We wish to perform level set-based topology optimization, but we also wish to ensure that the resulting shape is millable as defined in Section \ref{sec:constraint}.  We achieve this by incorporating millability into the definition of admissible shapes.  Thus in our setting, a shape is deemed admissible if the following definition holds.

\begin{defn}
\label{def:admissibility}
A shape $\Omega$ is \emph{admissible} if these requirements are met:
\begin{itemize}
	\itemsep = 0ex
	\item $\Omega$ is the interior region delimited by the zero contour of a piecewise smooth level set function; 
	\item $\partial \Omega$ contains all subsets where Dirichlet and Neumann boundary conditions from structural load cases are applied; 
	\item $\Omega$ wholly contained in a specified design domain; 
	\item $\partial \Omega$ is \emph{millable} according to Definition \ref{def:millability}.
\end{itemize}
\end{defn}

\subsection{The Archetypical Topology Optimization Problem for Millable Shapes}

The archetypical structural topology optimization problem, namely \emph{volume-constrained compliance minimization}, defines the objective function $\mathcal F$ as the average \emph{compliance} of $\Omega$ with respect to at least one load case, and imposes exactly one inequality constraint: namely that the volume of $\Omega$ must be less than or equal to some fraction of the volume of the design domain.  We will tackle this topology optimization problem in the present work, but with an additional millability constraint.  That is, we require the optimized shape to belong to the space of admissible shapes defined in Definition \ref{def:admissibility} above, ensuring that the optimized shape is millable as described in Section \ref{sec:millability}.  In other words, we solve the problem:
\begin{equation}
	\label{eqn:arch}
	\begin{aligned}
		\min_{\Omega \in \mathit{Adm}}  \qquad &\mathit{MeanCompliance}(\Omega) \\
		\mbox{s.t. } \qquad &\mathit{Volume}(\Omega) \leq V_0 
	\end{aligned}
\end{equation}
where $\Omega \in \mathit{Adm}$ if and only if $\Omega$ satisfies Definition \ref{def:admissibility}.  Here
$$\mathit{MeanCompliance}(\Omega) := \sum_{\ell=1}^L \frac{1}{L} \int_\Omega \sigma^{(\ell)}_\Omega : e^{(\ell)}_\Omega \, .$$
where $\sigma^{(\ell)}_\Omega$ and $e^{(\ell)}_\Omega$ are the stress and strain tensors in $\Omega$ induced by the $\ell^{\mathit{th}}$ structural load case.

\subsection{Maintaining Admissibility in Level-Set Based Topology Optimization}
\label{sec:millability}

Thanks to the admissibility criteria of Definition \ref{def:admissibility}, the algorithm outlined in the Section \ref{sec:to} must be modified slightly before it can be applied to solve the optimization problem \eqref{eqn:arch}.  This is because the choice of speed function \eqref{eqn:speedchoice} is no longer suitable as it can violate admissibility.  For example, suppose that advection with respect to $- d \mathcal L $ causes a part of the boundary of the shape at the next iteration to become inaccessible with respect to the specified milling direction through occlusion with another part of the boundary of the shape.  To rectify this issue, we propose in this paper a strategy for modifying the speed function \eqref{eqn:speedchoice} so that the shape remains millable from one iteration to the next while still ensuring that the Lagrangian decreases.

We are inspired by the classical \emph{gradient projection method} \cite{nocedal} in which the outward-pointing components of the descent direction orthogonal to the active constraint boundary are set to zero in order to prevent violation of the constraints in those components.  We adapt this method to level set-based shape optimization as follows.   We first assume that our initial shape does not violate the millability condition, which is easily achieved by initializing with a convex hull of the input geometry for example.  Next, in each iteration we modify the speed function at each point of the boundary of the shape by smoothly transitioning it to zero when a violation of millability is predicted in the next iteration.  The goal is to force the boundary velocity to vanish where a violation of millability is predicted, thereby halting any movement of the boundary there.

To be precise, we introduce a ``filter" function $\filter$ (defined precisely in Section \ref{sec:filter} below) that assigns a value in $[0,1]$ to each point of the boundary.  It assigns the value zero to any point belonging to the subset $\Gamma \subseteq \partial \Omega$ where a violation of millability is detected. Its values then transition smoothly from 0 to 1 in a small ``collar" around the periphery of $\Gamma$, and it assigns the value one to all other points of $\partial \Omega$.  To obtain the modified speed function, we \emph{multiply} the original speed function by $\filter$.  In other words, 
$$ v_{\mathit{modified}} := - \eta d\mathcal L \, .$$
It is important to realize that, since $\filter$ is everywhere positive, then the modified boundary speed remains a descent direction for the Lagrangian.  This is easily seen by substituting $v_{\mathit{modified}}$ for $\speed$ in formula \eqref{eqn:descent}.  Therefore we reduce the value of the Lagrangian during each iteration of topology optimization while only allowing the shape to be modified from directions accessible by the milling tool during that iteration.  

\subsection{Definition of the Filter}
\label{sec:filter}

We now define the function $\eta$ precisely.  Note that the velocity of a point $\surfacePt \in \partial \Omega$ has the form $\velocity(\surfacePt) = \speed(\surfacePt) \normal(\surfacePt)$, where $\normal(\surfacePt)$ is the outward-pointing unit normal vector field at $\surfacePt$.  If $\speed(\surfacePt) >0$ then the boundary of the shape grows near $\surfacePt$; if $\speed(\surfacePt) <0$ then the boundary of the our first requirement for the definition of $\eta$ is that $\eta(x) v(x) \leq 0$ for all $x$ on the boundary of the shape, i.e.\ preventing growth, as this is a means to avoid occlusion of previously accessible regions of the surface of the shape.  Next we choose the ``best'' milling direction from amongst all ``accessible'' ones, where a direction is accessible if there is no intersection between the shape and the tool bit or head (see Figure \ref{fig:constraintEx}), and best means closest to the surface normal. 

To state the above considerations rigorously, let $V_i(x)$ denote the volume occupied by the union of the milling tool and head when it is oriented along the milling direction $\millingDir_i$ and the tip of the head is located at $x$.  Then we define $\filter$ as follows:
\begin{equation}
\label{eqn:filter}
\filter(x) = 
\begin{cases}
0 &\quad \text{if } \speed(x) \geq 0 \\
\max\{ {|\millingDir_i \cdot \normal(x)|} : \mbox{all } \millingDir_i \in \millingDirSet \mbox{ s.t. } \millVolume_i(x) \cap \Omega = \varnothing \} &\quad \text{otherwise.}
\end{cases}
\end{equation}
The initial set of milling directions $\millingDirSet$ is either specified by the user depending on their desired machine configuration; or for 5-axis milling, we explore several methods in the following section. 

\begin{figure}[h]
	\begin{center}
		\small
		\includegraphics[width=150mm]{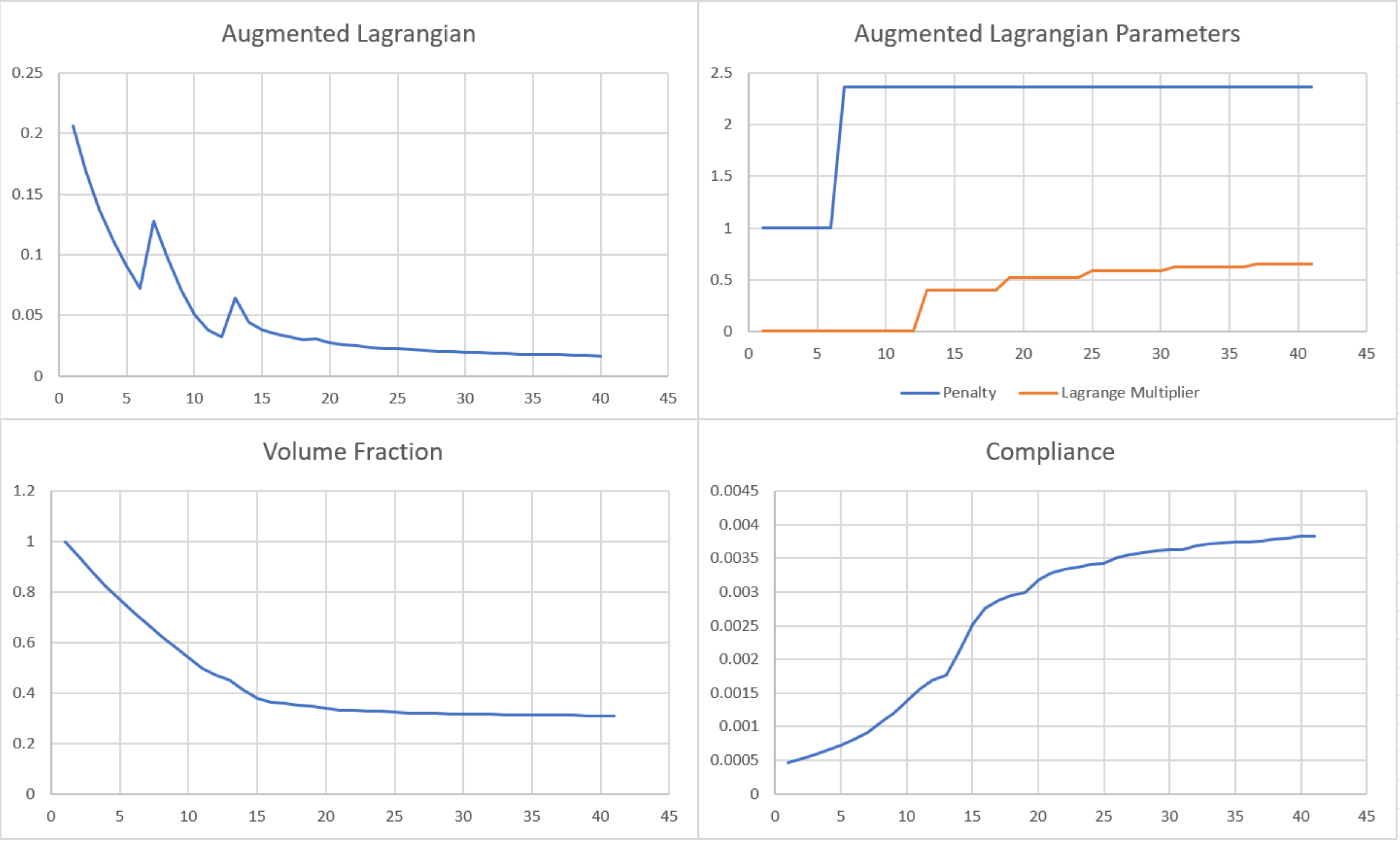}  
		\caption{Optimization algorithm data as a function of iteration for the {\bf TorqueStruct} example (see Section \ref{sec:PDs} for problem definition). The relative rate of change of the augmented Lagrangian and the volume fraction constraint violation in the final iteration are approaching $1\%$.}
		\label{fig:cvgce}
	\end{center}
\end{figure}

\subsection{Convergence Properties}

The fact that our algorithm chooses a descent direction for the augmented Lagrangian means we can ensure that the value of the augmented Lagrangian decreases in every iteration of each inner loop of the augmented Lagrangian algorithm (assuming the advection time is chosen with a backtracking line search, as we do).  This fact alone certainly does not guarantee the convergence of the sequence of shapes to a locally optimal shape satisfying the constraints; and a rigorous analysis of the convergence properties of our algorithm is beyond the scope of this paper (and would undoubtedly be quite difficult).  However, we do observe in all our examples that the norm of the filtered shape gradient of the augmented Lagrangian decreases to zero within acceptable tolerances in each inner loop of the augmented Lagrangian algorithm.  We also observe that the volume fraction constraint violation decreases to zero within acceptable tolerances over the course of the algorithm, and the millability constraint is always satisfied by construction.  See Figure \ref{fig:cvgce} for a graph of the relevant data for the augmented Lagrangian algorithm as a function of iteration (i.e.\ augmented Lagrangian value, volume fraction, compliance, Lagrange multiplier, penalty ceofficient) typical of our algorithm.

It is well-known that gradient-based optimization algorithms applied to highly non-convex problems such as topology optimization have a  tendency to get stuck in local minima.  Our algorithm is no exception.  In fact, it may perform less well than conventional level set-based topology optimization from this point of view because it is likely that the ``projected'' nature of our algorithm may cause it to stop prematurely at an undesirable locally optimal shape for which there exists a nearby millable shape with better performance that could easily be reached using an un-filtered descent step but not a filtered descent step.  This is a limitation of our algorithm.  However, the shapes that we can reach using our algorithm are nevertheless quite acceptable.

\section{Implementation}
\label{sec:impl}

The interesting implementation challenge is to compute for any choice of $i$ and $x$ the sets $\millVolume_i(x) \cap \partDomain$ appearing in the milling constraint defined in Section \ref{sec:algorithm}. Conveniently, we can use a level set representation for $\millVolume_i(x)$ and then take advantage of a collection of level set operations that together lead to a very efficient algorithm.  In addition this algorithm can be run in parallel to take advantage of modern multi-core architectures. 

We implemented the volume-constrained compliance minimization problem using the level set-based topology optimization method outlined in $\cite{Alla04}$ but with the modification of the boundary speed function described in the previous section.  We used the level set library from $\cite{Museth13}$ that creates signed distance functions and has implementations for the following functions: ray-casting, computing offsets, advection, and morphological closure.  For completeness, these are given as Algorithms 1 through 4 in the Appendix.

\subsection{The Millability Filter in the Three-Axis Case}

\begin{figure}[h]
	\begin{center}
		\small
		\includegraphics[width=73mm]{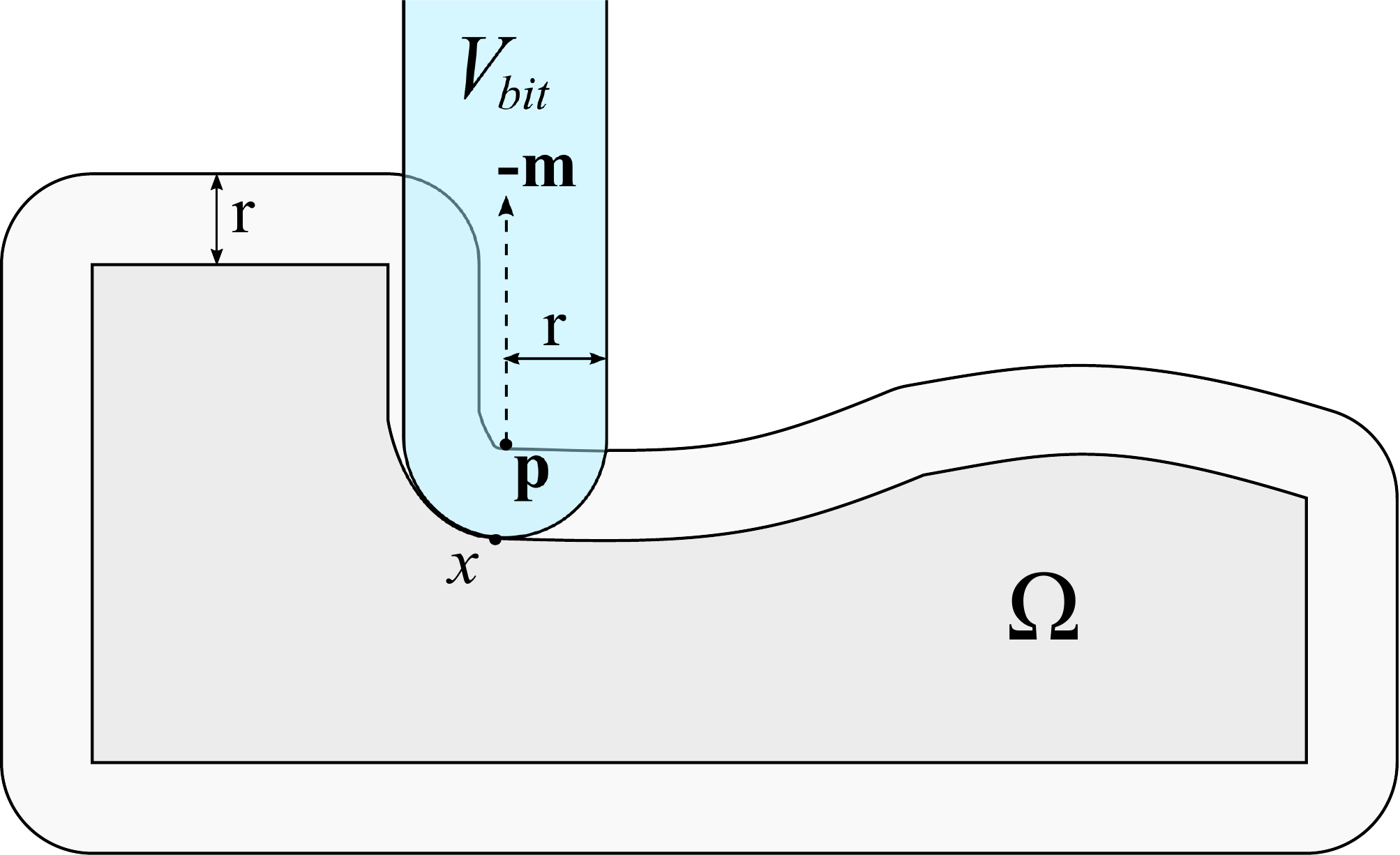}
		\caption{Example showing that a ray casting test from $\toolTip$ in $-\millingDir$ against $\partDomain$ offset by $\bitradius$ is equivalent to testing whether $\millVolume_{bit} \cap \partDomain \neq \varnothing$ }
		\label{fig:int}
	\end{center}
\end{figure}

The algorithm used to compute $\filter$ for 3-axis machines is as follows: We find a set of points $\surfacePt$ of the boundary of $\partDomain$ by projecting grid points in the narrow-band of $\sdf$ onto the zero level set along $\sdfGrad$. Then for each $\surfacePt$, we loop over each milling direction and test for accessibility.  The accessibility test combines two ray casting operations. The first tests against the bit intersection with the part and the second tests against mill head intersection with the part. One can see from Figure \ref{fig:int} that a ray originating at $\toolTip$ in direction $-\millingDir$ will only intersect the iso-surface of $\partDomain$ offset by $\bitradius$ if the bit will intersect with the part (i.e.\ $\millVolume_{bit} \cap \partDomain \neq \varnothing$ where $\millVolume_{bit}$ denotes the volume occupied by the bit).  A similar test is also performed for the mill head except the iso-surface at $\headradiusA$ is used for the ray cast instead and the ray origin is $\toolTip - \millingDir_i(\bitlength+\headradiusA)$.  Combining these two tests we can easily determine $\millVolume_i(x) \cap \partDomain$. Note that when using a narrow-band representation for $\partDomain$, the narrow-band must be extended to include iso-surfaces offset up to $\headradiusA$. This process can be performed once during initialization using a standard level set offset function. Pseudo code for the above can be found in the appendix as Algorithm \ref{alg:filter} and Algorithm \ref{alg:millTest}.

\subsection{The Millability Filter in the Five-Axis Case}

In the following sections we explore several different methods for obtaining the most accessible milling direction for a 5-axis machine. All of the following methods begin by testing the milling direction opposite to the surface normal $\millingDir_0=-\normal$. This direction would provide the maximal value for $\filter(x)$. But it is possible for the surface point $x$ being tested to be inaccessible from this direction, yet accessible from another direction.  To find another direction, we propose several methods.  All our methods are essentially searches on the hemisphere above $x$, with ever-more refined search strategies.

\begin{figure}[h]
	\begin{center}
		\small
		\includegraphics[width=50mm]{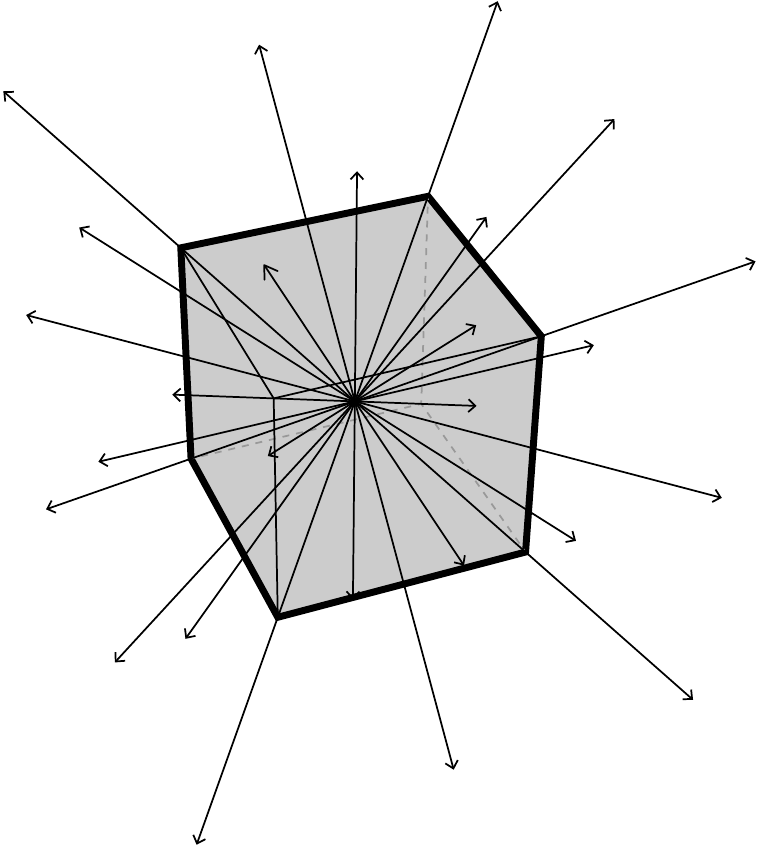}
		\caption{The set of milling directions $\millingDirSet$ used in our implementation of `Hemisphere sampling'. }
		\label{fig:cubeDirs}
	\end{center}
\end{figure}

\paragraph{Hemisphere sampling.}
In this method we subsample the hemisphere to obtain a discrete set of directions $\millingDirSet$ and test each to find the most accessible direction. In our implementation we used a fixed set of 26 samples on the sphere to form $\millingDirSet$. If we consider the center of a cube, the milling directions point from the cube center to the center of every face, the midpoints of all the edges and the cube corners (See Figure \ref{fig:cubeDirs}). These directions were all normalized. We used the 3-axis algorithm (Algorithm \ref{alg:filter}) with this $\millingDirSet$ to approximate a 5-axis procedure, and we note that in this procedure the accessibility test automatically filters $\millingDirSet$ to those in the hemisphere such that $\normal(\surfacePt) \cdot \millingDir_i < 0$ (Algorithm \ref{alg:millTest}).

This method has several drawbacks: first, all directions in $\millingDirSet$ must be tested to determine the most accessible, second it is possible that all of the directions are inaccessible but a small perturbation of one of the directions could be actually accessible. The advantage of this method is that it does not just perform a local search of part of the sphere and will not be trapped by concavities. 

\begin{figure}[h]
	\begin{center}
		\small
		\includegraphics[width=60mm]{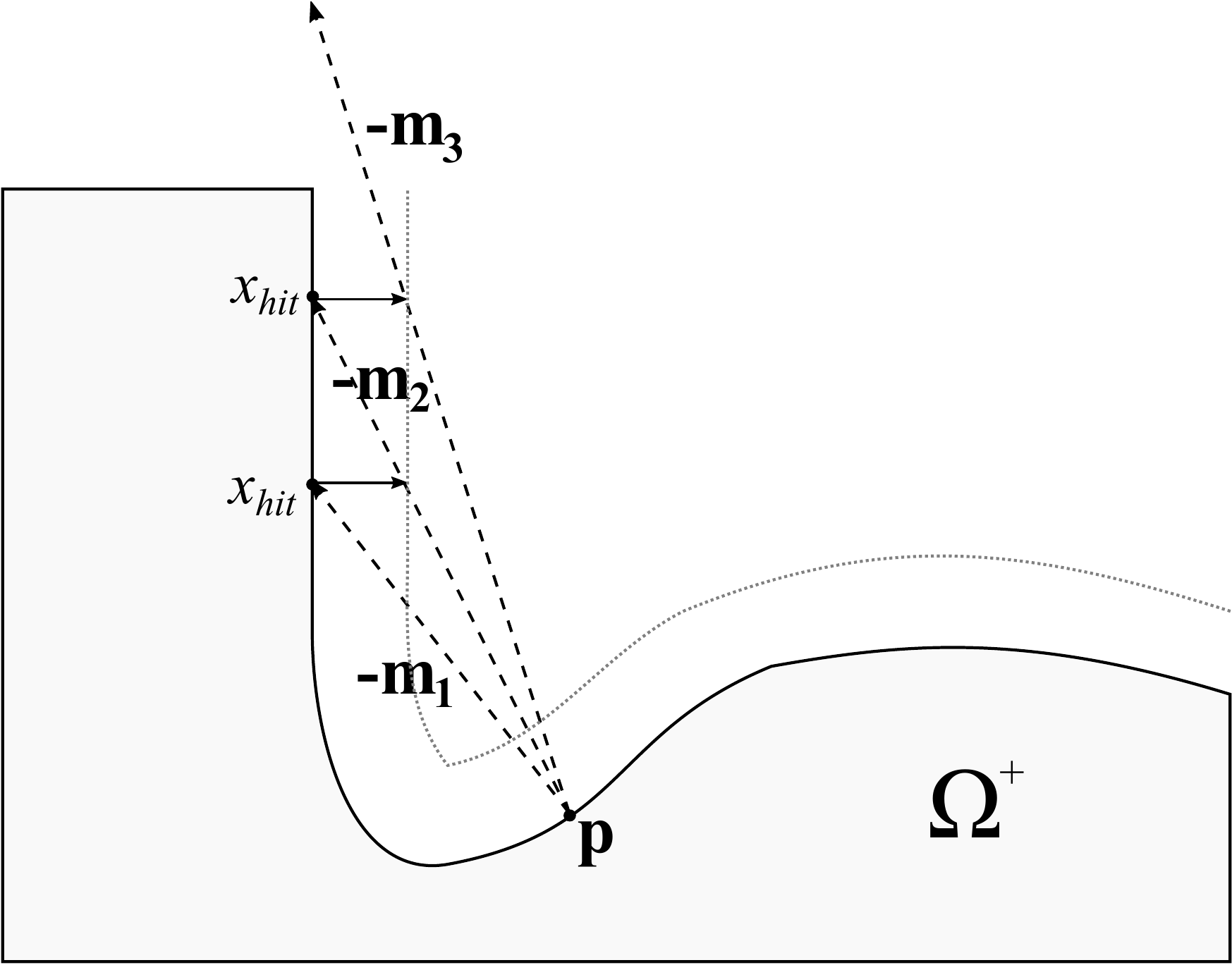}
		\caption{Example showing a successful result from the `Normal Search' method for determining an accessible milling direction. The search algorithm begins with $\millingDir_1 = -\normal$ and performs a ray-trace on $\partDomain^+$, which is $\partDomain$ offset by the bit or head radius. The intersection $\intPt$ is found and a new milling direction $\millingDir_2$ is found by offsetting from $\partDomain^+$ in the normal direction up to the periphery of the narrow-band (shown as dotted gray line). Direction $\millingDir_2$ is then tested with another ray-trace, and in the same way leads to $\millingDir_3$ which finally results in no intersection with $\partDomain^+$ and this direction is returned as a candidate milling direction.}
		\label{fig:normalSearch1}
	\end{center}
\end{figure}

\begin{figure}[h]
	\begin{center}
		\small
		\includegraphics[width=60mm]{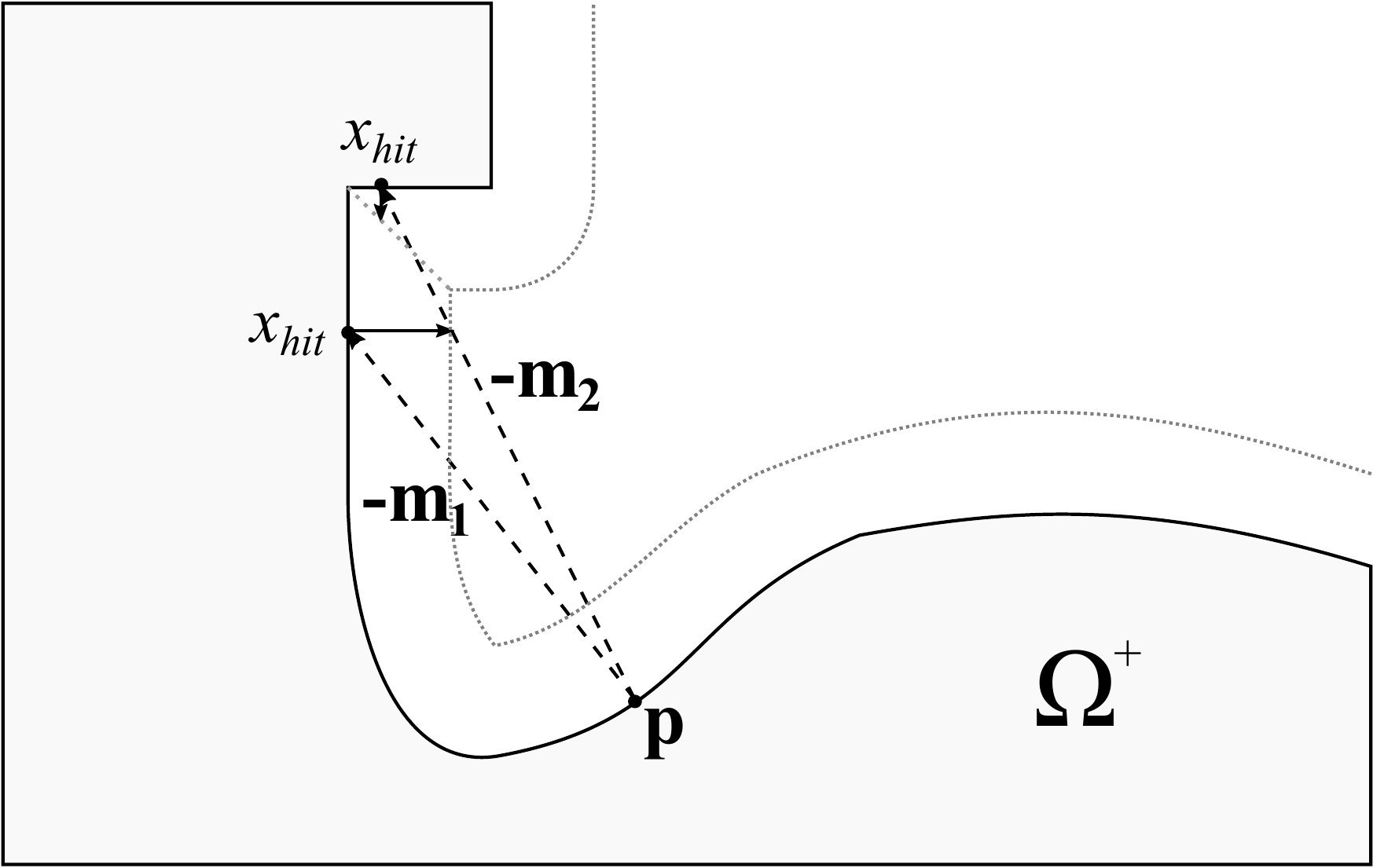}
		\includegraphics[width=60mm]{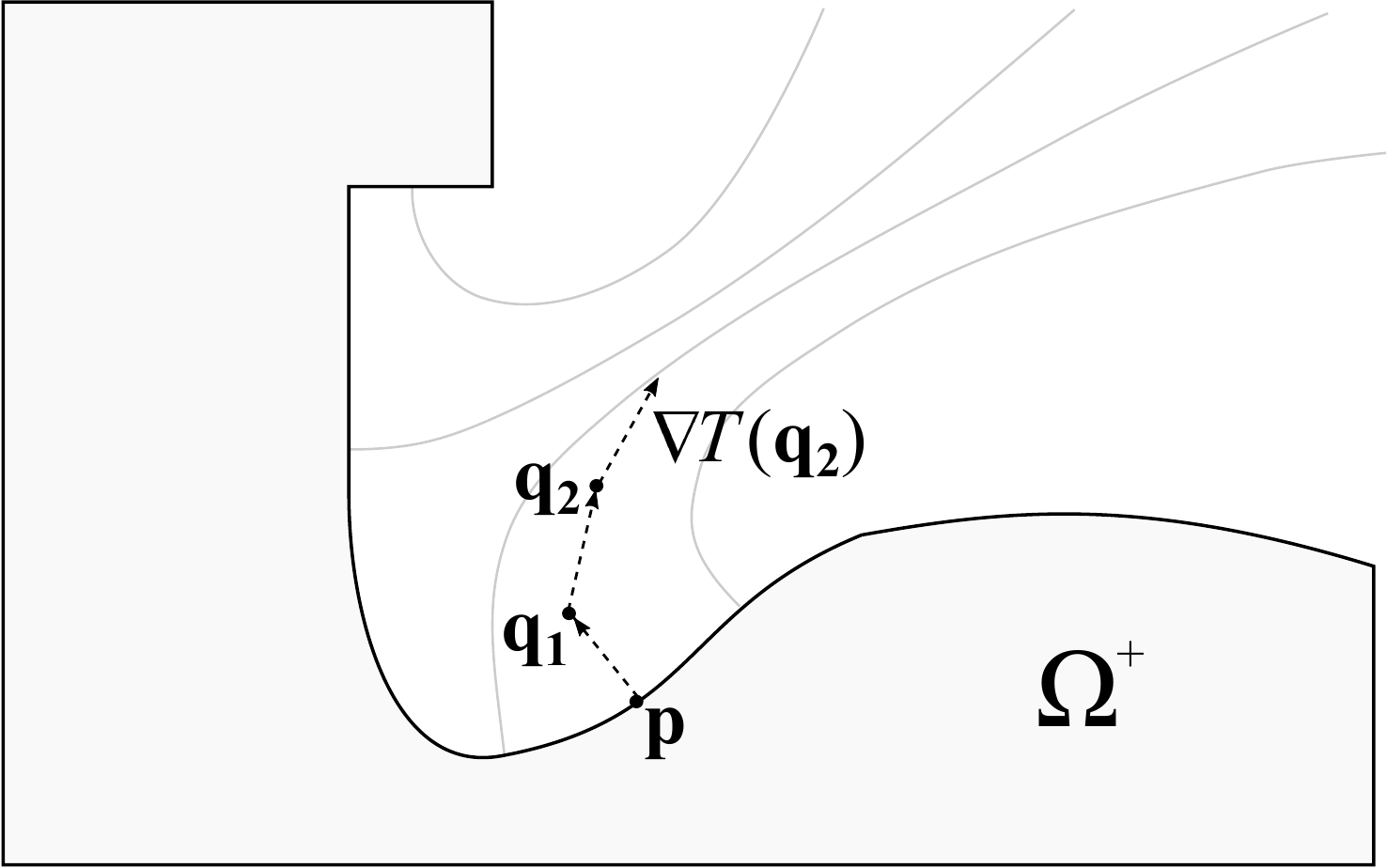}
		\caption{{\bf Left:} Example showing an unsuccessful result from the `Normal Search' method for determining an accessible milling direction and {\bf Right:} a successful result from `Heat Search'. The `Normal Search' follows milling directions $-\millingDir_1$, intersects at $\intPt$ and offsets to the narrow-band to find $-\millingDir_2$. When following $-\millingDir_2$ a new intersection is found but the new offset does not escape from the overhanging geometry and is unable to find an accessible direction even though one exists. The `Heat Search' follows a trajectory through the gradient of the heat $\TempFieldGrad(\heatPt)$ which quickly finds an accessible direction. Other trajectories are shown in light gray illustrate $\TempFieldGrad$.}
		\label{fig:normalSearch2}
	\end{center}
\end{figure}

\paragraph{Normal search.}
A dense sampling of the hemisphere is not practical so in this method we rely on the signed distance field $\sdf$ around $\partDomain$ to guide a local search of the hemisphere. If a milling direction is tested and fails, the intersection hit-point $\hitPt$ from the ray-trace, along with an offset is used to determine a new milling direction. This is done by moving from $\hitPt$ in the direction of $\sdfGrad(\hitPt)$ until either the medial axis of $\partial \Omega$ or the periphery of the level set narrow band is reached --- determined  by finding either a peak or a plateau along the line emanating from $\hitPt$ in the direction $\sdfGrad(\hitPt)$, respectively.  A peak is a point $y$ where the signed distance ceases increasing beyond $y$ in the search direction, while a plateau consists of points $y$ such that $\sdf$ is constant near $y$.  This new direction is in turn tested and updated until an accessible direction is found or an iteration limit is reached (see Figure \ref{fig:normalSearch1}). The advantage of this method is that it is a local search and often will quickly find an accessible direction in very few iterations (typically fewer than a coarse sampling of the whole hemisphere). The disadvantage of this method is that it is not guaranteed to find a solution even if one exists. It is possible for the method to move away from the accessible direction in some corner cases (see Figure \ref{fig:normalSearch2}). Details are shown in Algorithm \ref{alg:normalSearch}.
\bigskip

\begin{figure}[h]
	\begin{center}
		\small
		\includegraphics[height=65mm]{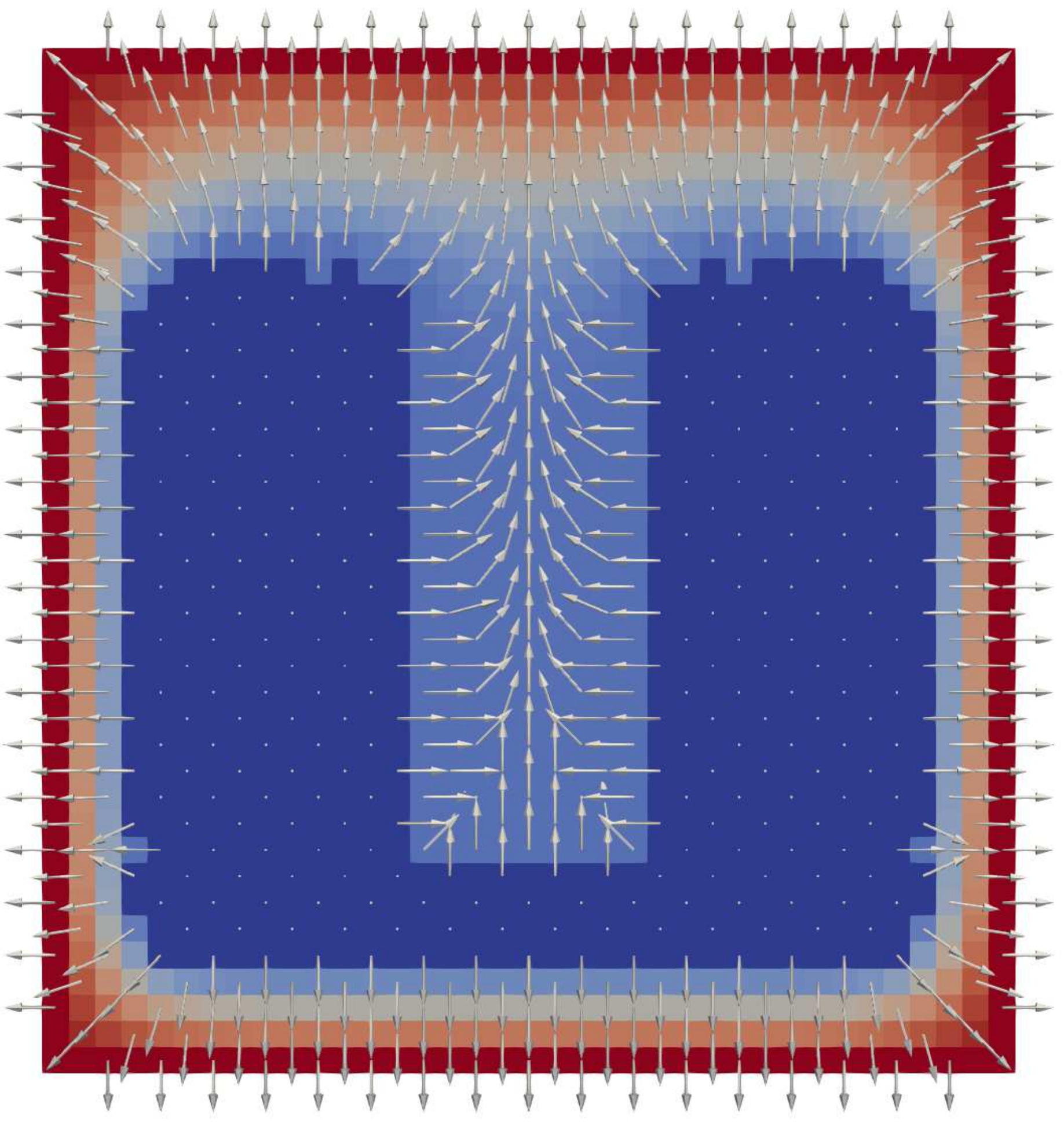}
		\caption{Example showing the heat field $\TempField$ for a slice through the center of {\bf TorqueStruct} (see Figure \ref{fig:supportPD} right). An arrow plot illustrates the gradient of the heat field $\TempFieldGrad$ and shows how trajectories in this field lead from the surface of $\partDomain^+$ (blue) to the accessible boundary (red). }
		\label{fig:concavityHeat}
	\end{center}
\end{figure}

\paragraph{Heat search.}
This method is an extension to the `Normal search' method that handles the drawback of moving away from accessible directions. Instead of following the surface normal at the intersection point, we follow the direction of heat diffusion away from the surface toward the bounding box of the geometry. This requires an initial step of solving the transient heat equation to an approximate steady-state in the volume between the surface of $\partDomain^+$ ($\partDomain$ offset by $\bitradius$) and the bounding box \cite{Widd75}. For this, we voxelize this domain and use a finite difference method to discretize the spatial derivatives as well as a first-order Euler method with sufficiently small time-step $\heatEps$ to discretize the time derivatives.  We set two Dirichlet boundary conditions: on the vertices of the boundary of the voxel grid adjacent to $\partDomain^+$ we set the temperature to zero; and on the vertices of the bounding box we set the temperature to one. Once we have the temperature field $\TempField$ on the voxel grid, we integrate trajectories tangent to $\TempFieldGrad$ using sub-voxel multi-linear interpolation of $\TempField$ together with a first-oder Euler method.  Thus the gradient of the temperature field guides the search for the milling direction (See Figure \ref{fig:concavityHeat}). 
Note that it is important to solve the heat equation between the offset surface $\partDomain^+$ and the bounding box and not $\partDomain$ and the bounding box since the offset surface will close off some exits from cavities that are narrower than the bit diameter.
\bigskip

\subsection{Finite Element Analysis}

In order to compute the mean compliance of a shape $\Omega$ and its shape gradient, we must solve the linear elasto-static equations in $\Omega$ with respect to one or more load cases.   To this end, we use a finite element-based structural solver (Autodesk NASTRAN) in a discretization of $\Omega$ into elements according to the following procedure. 

The level set function defining the shape at each iteration is defined on a uniform, background Cartesian grid.  We do not use this grid for the the structural solver.  Instead, in the first iteration we define a co-located conforming tetrahedral mesh using the meshing capabilities of the solver (essentially this performs a standard level set to mesh conversion on the level set function defining the initial shape). In the first iteration, we use this mesh to compute the finite element solution.  In subsequent iterations, when the shape no longer conforms to this mesh, we nevertheless use an ``active" subset of this mesh for the finite element solution.  To be precise, we use the values of the level set function to detect which tetrahedra are fully inside the shape, which are cut by the boundary of the shape, and which are fully outside the shape.  We then solve the discrete linear elasto-static equations only in the interior and cut tetrahedra. The stiffness matrix for the union of these tetrahedra is assembled from the interior tetrahedra in the usual way, and by multiplying the contribution from each cut tetrahedron with its volume fraction of intersection with the shape (and thresholded by a small value to avoid ill-conditioning).  We use linear Lagrange shape functions inside these elements.

The load vectors corresponding to each load case can always be computed because the surface patches where the surface loads and prescribed displacements of each load case are applied belong to the evolving shape at every iteration. This is because we set the speed function governing the shape updates to zero on these surface patches.  Therefore, since the initial mesh contains a triangulation of these surface patches, the subset of the mesh used for the finite element solution at every subsequent iteration does as well.

Once the solutions to the discrete linear elasto-static equations have been computed, we construct their per-element strain energy densities (the negative shape gradient of the compliance is the strain energy density) in the boundary tetrahedra of the active subset of the mesh.  The negative shape gradient of the augmented Lagrangian at the current iteration is a linear combination of the sum of the strain energy densities and the negative shape gradient of the volume (equal to $-1$). We assign to the grid nodes nearest the boundary of the shape at the current iteration the value of the negative shape gradient in the nearest boundary tetrahedron.  We then extend these values to a narrow band of the boundary of the shape using the Hamilton-Jacobi equation for normal extension described in Section \ref{sec:algorithm}. 

\subsection{Shape Update Strategies}

Finally we conclude with some implementation details concerning the way in which the shape is updated and how the milling constraint fits into it.  We have two variants: an algorithm that assumes initial millability and ensures millability in each subsequent iteration as described in Section \ref{sec:algorithm}; and an algorithm that initially relaxes the millability constraint and enforces it only at convergence.

\paragraph{Strict algorithm.}
At each iteration the negative shape gradient of the augmented Lagrangian is evaluated with finite element analysis and extended to a narrow band of the shape boundary as described in the previous section.  We obtain the advection speed function by multiplying the extended negative shape gradient with the milling filter $\filter$ as described in Section \ref{sec:algorithm}.  Positive speeds that could allow for shape growth are removed in this way.  The shape $\partDomain$ is then updated by level set advection with this speed function. This algorithm ensures that we only remove accessible material at each step and it requires that the initial shape of $\partDomain$ be fully accessible (a feasible start). Convergence can be tested when either the speed approaches zero or the change in objective function is below a threshold. The final step performs the morphological closing of $\partDomain$, which explicitly enforces fillets and fills any cavities smaller than the tool radius $\bitradius$ that have been created. The result of this operation is a very minor cleaning of the level set since the fillets and curvature of cavities are already implicitly controlled by the filter function $\filter$. See the details in Algorithm \ref{alg:strictAlgorithm}.

\paragraph{Relaxed algorithm.}
A more relaxed version of the above algorithm allows for positive values in the speed function so that the algorithm can begin with an infeasible starting condition before eventually converging to a feasible solution. In order to converge to a feasible solution we apply a positive velocity whenever we detect inaccessible regions (i.e. points where $\filter = 0$) thus filling in cavities or otherwise inaccessible regions of $\partDomain$. Values between $\posVal=0.1$ and $\posVal=0.25$ work well although there is a trade-off between speed of correcting inaccessible regions with interfering with the subtractive milling. When $\posVal$ is too large, surface regions that were inaccessible may grow too much and this overshooting will need to be compensated with material removal in subsequent iterations. See Algorithm \ref{alg:relaxedAlgorithm} for details.

\section{Results}

\subsection{Problem setups}
\label{sec:PDs}

\begin{figure}[h!]
	\begin{center}
		\small
		\includegraphics[height=45mm]{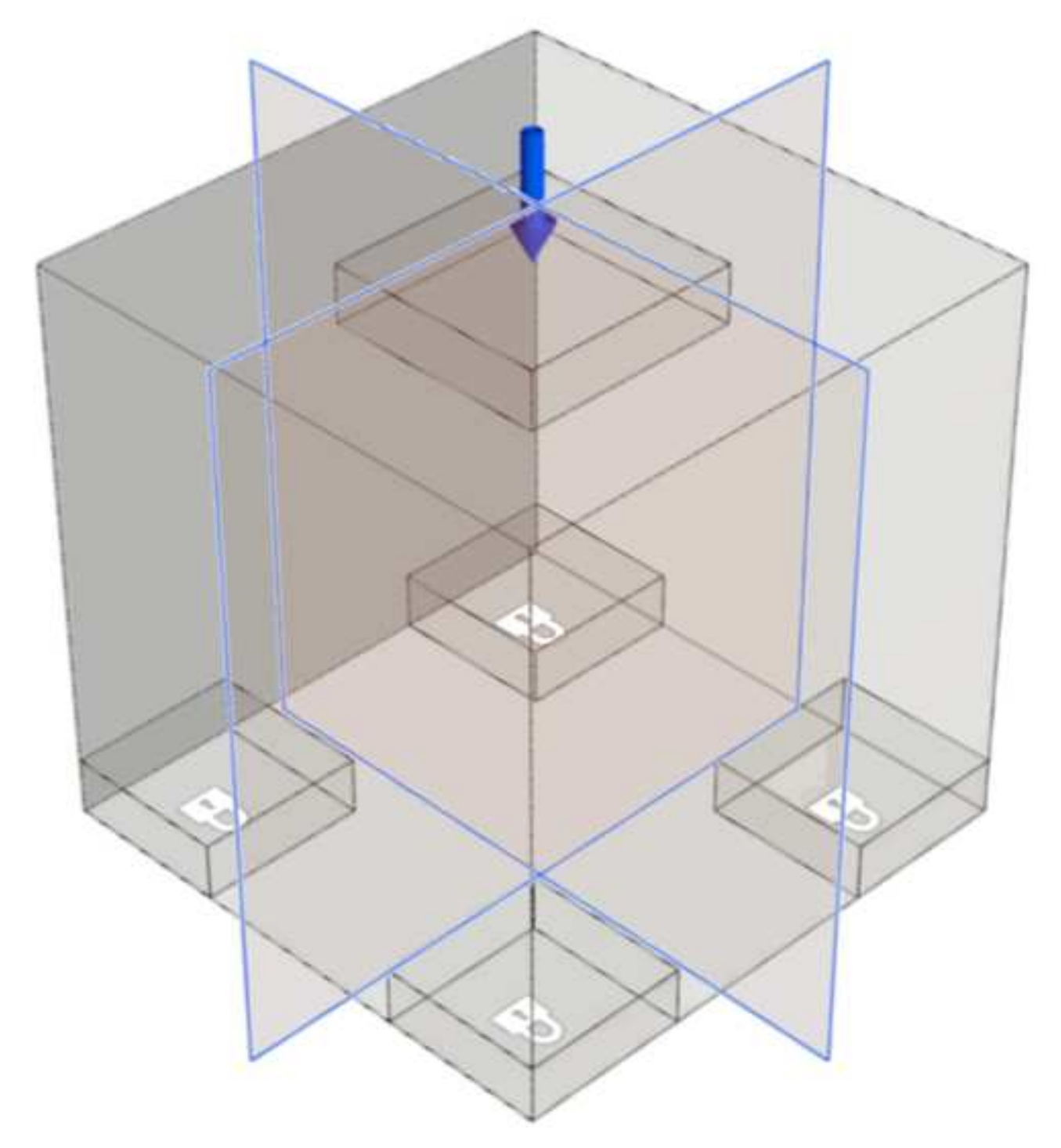}
		\includegraphics[height=45mm]{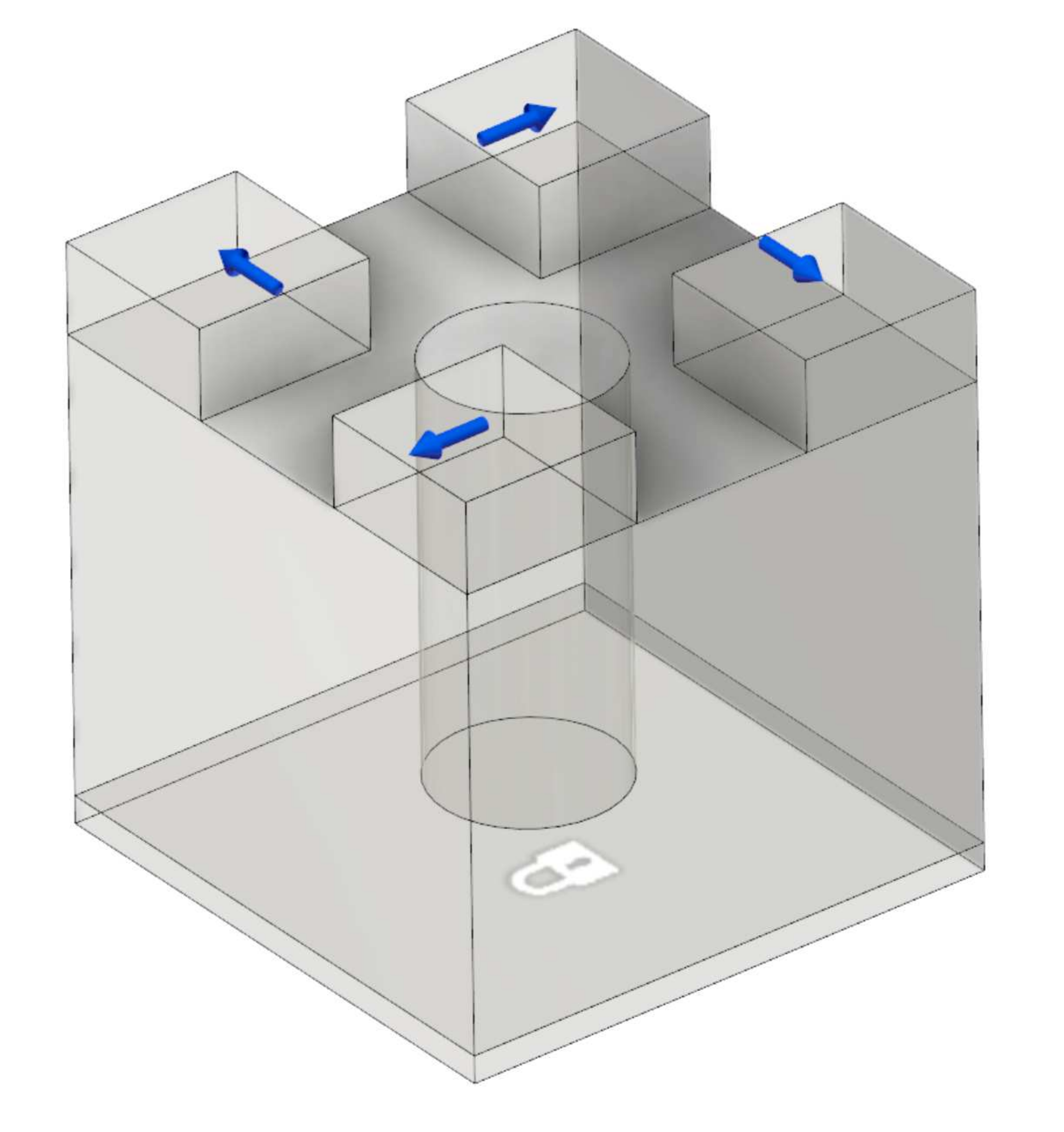}
		\caption{{\bf Left}: The problem setup for the {\bf SupportStruct}: The design space is a box enclosing the top plate that is loaded from above and supported by four plates below with Dirichlet boundary conditions on their bottom faces (indicated by the lock symbols). Two symmetry planes cut through the design space. {\bf Right}: The problem setup for the {\bf TorqueStruct}: The design space is the union of the shown boxes with a central cylindrical column subtracted away. The single load case is shown where the top plates are loaded with a twisting type load. The bottom plate is fixed with a Dirichlet boundary conditions on its bottom face. }
		\label{fig:supportPD}
	\end{center}
\end{figure}

\paragraph{SupportStruct.} This problem setup consists of a plate supported from four foot plates (See Figure \ref{fig:supportPD}). We used a single load case with a vertical mechanical load of 3KN applied to the top surface of the top plate, while the four lower plates have fixed Dirichlet boundary conditions on their bottom surfaces. The plates and material are modelled as uniform in density with a Poisson ratio of 0.29 and Young's modulus of 17 GPa. In this problem we also define a pair of symmetry planes and enforce mirror symmetry in the output geometry across the planes. The dimensions of the bounding box of the geometry are 50mm $\times$ 50mm $\times$ 50mm.

\paragraph{TorqueStruct.} This problem setup consists of a set of four plates supported with a set of four loads creating a twisting load and supported by a fixed plate below (See Figure \ref{fig:supportPD}). We also use a single load case and each loaded plate receives a mechanical load of 50N applied to its top surface and the lower plate has a fixed Dirichlet boundary condition on its bottom surface. The plates and material are modelled as uniform in density with a Poisson ratio of 0.3 and Young's modulus of 210 GPa. The dimensions of the bounding box of the geometry is 60mm $\times$ 63mm $\times$ 60mm.

\begin{figure}[h!]
	\begin{center}
		\small
		\includegraphics[height=45mm]{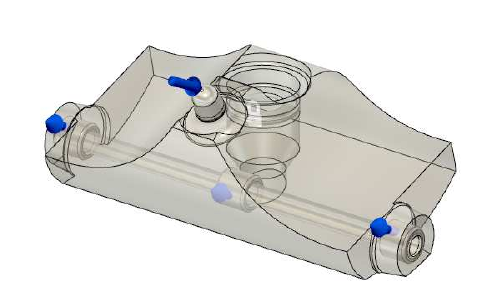}
		\includegraphics[height=45mm]{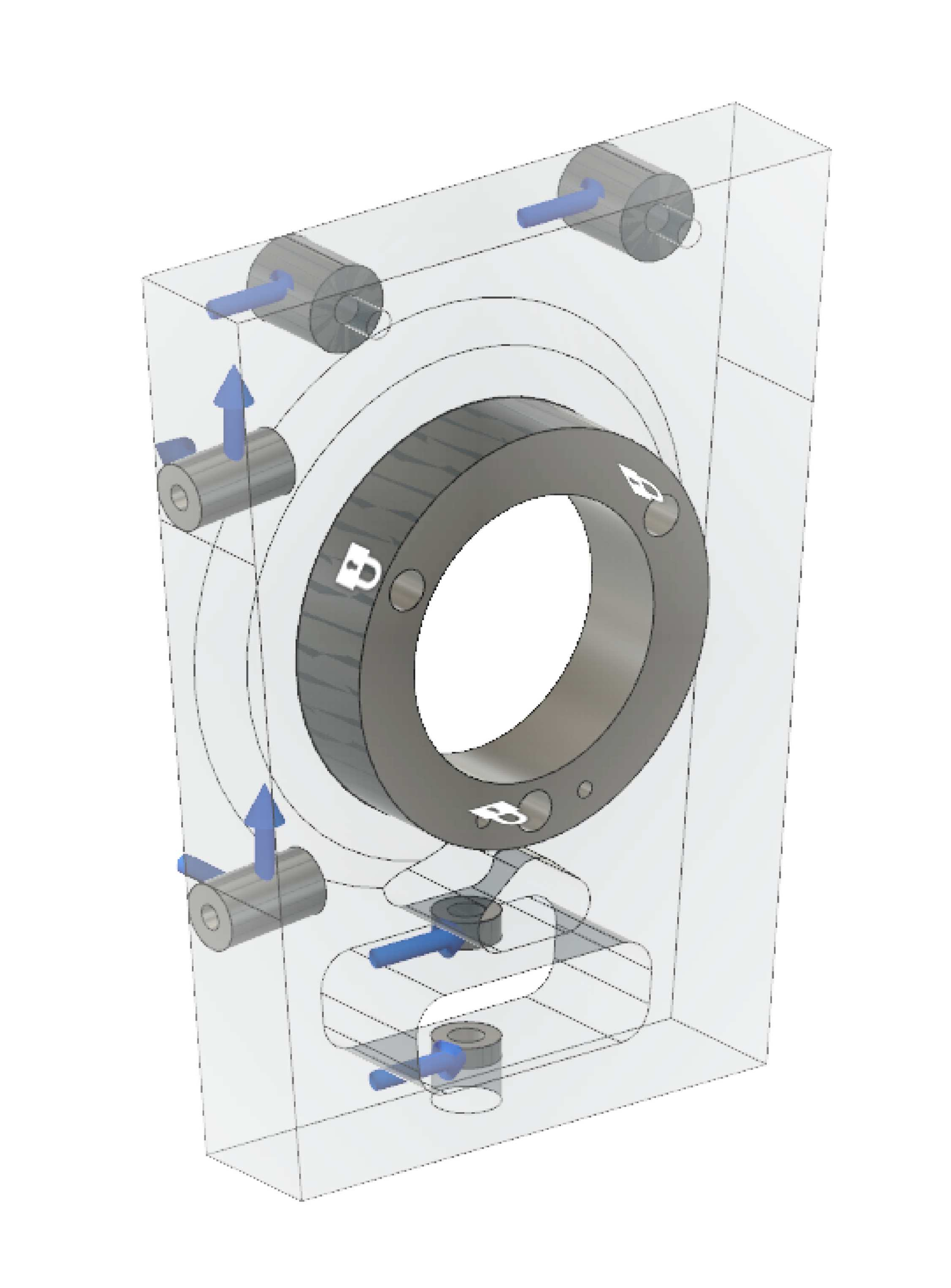}
		\caption{One of the load cases used to define the problem along with the design space for {\bf Left}: {\bf SkateTruck}. {\bf Right}: {\bf Upright}.}
		\label{fig:skatePD}
	\end{center}
\end{figure}

\paragraph{SkateTruck.} This problem setup consists of the support structure for the axle and connecting to the body of a skateboard. This case is provided as an illustration for the subtractive constraint applied to a more complex real-world problem. It consists of 6 load cases, one is illustrated in Figure \ref{fig:skatePD}. The material is Aluminum, modelled as uniform in density with a Poisson ratio of 0.33 and Young's modulus of 68.9 GPa.  The dimensions of the bounding box of the geometry are 188mm $\times$ 44mm $\times$ 151mm.

\paragraph{Upright.} This problem setup consists of the upright in a suspension system of a performance automobile. This case is also provided as an illustration of the constraint applied to a more complex real-world problem. One of the 5 load cases is illustrated in Figure \ref{fig:skatePD}. The material is Aluminum, modelled as uniform in density with a Poisson ratio of 0.33 and Young's modulus of 68.9 GPa.  The dimensions of the bounding box of the geometry are 188mm $\times$ 44mm $\times$ 151mm.

\subsection{Optimization algorithm}
\label{sec:optAlg}
In our experiments on the {\bf SupportStruct} and {\bf TorqueStruct} we ran level set topology optimization using the Augmented Lagrangian method described in Section \ref{sec:algorithm} but with boundary speed modification as described in Section \ref{sec:algorithm}.  We minimize compliance subject to a volume fraction constraint of 20 and 30 percent of the initial design space respectively. 

In our experiments on the {\bf SkateTruck} and {\bf Upright} we ran level set topology optimization using a proprietary method, with the objective of minimizing compliance subject to a volume fraction constraint of 30 percent of the initial design space, but now including a stress constraint. The proprietary aspect of the algorithm is how the stress constraints are handled.  Our intention with this experiment is to highlight the versatility of our millability filtering approach --- it can be applied to any shape update scheme to produce millable results.  Indeed, the proprietary method assembles the boundary speed from the shape gradients of the objective and constraint functions in a different manner than the ``standard" method described in Section \ref{sec:algorithm}.  Nevertheless, when we modify this algorithm's boundary speed exactly as described in Section \ref{sec:algorithm}, we obtain the expected outcome that the Lagrangian still decreases for sufficiently small shape updates while the millability of the updated shape is maintained.

In all the experiments, the geometry with loads and fixity applied are all considered preserved regions and at each step in the optimization forced to remain part of the optimized geometry via boolean union operations. In all the examples, the relaxed algorithm was used to allow for recovery from infeasible states with $\posVal=0.25$.

\subsection{Parameter experiments}

\begin{figure}[h]
	\begin{center}
		\small
		\begin{tabular}{ m{5cm} m{5cm} m{5cm} }
			\includegraphics[width=40mm]{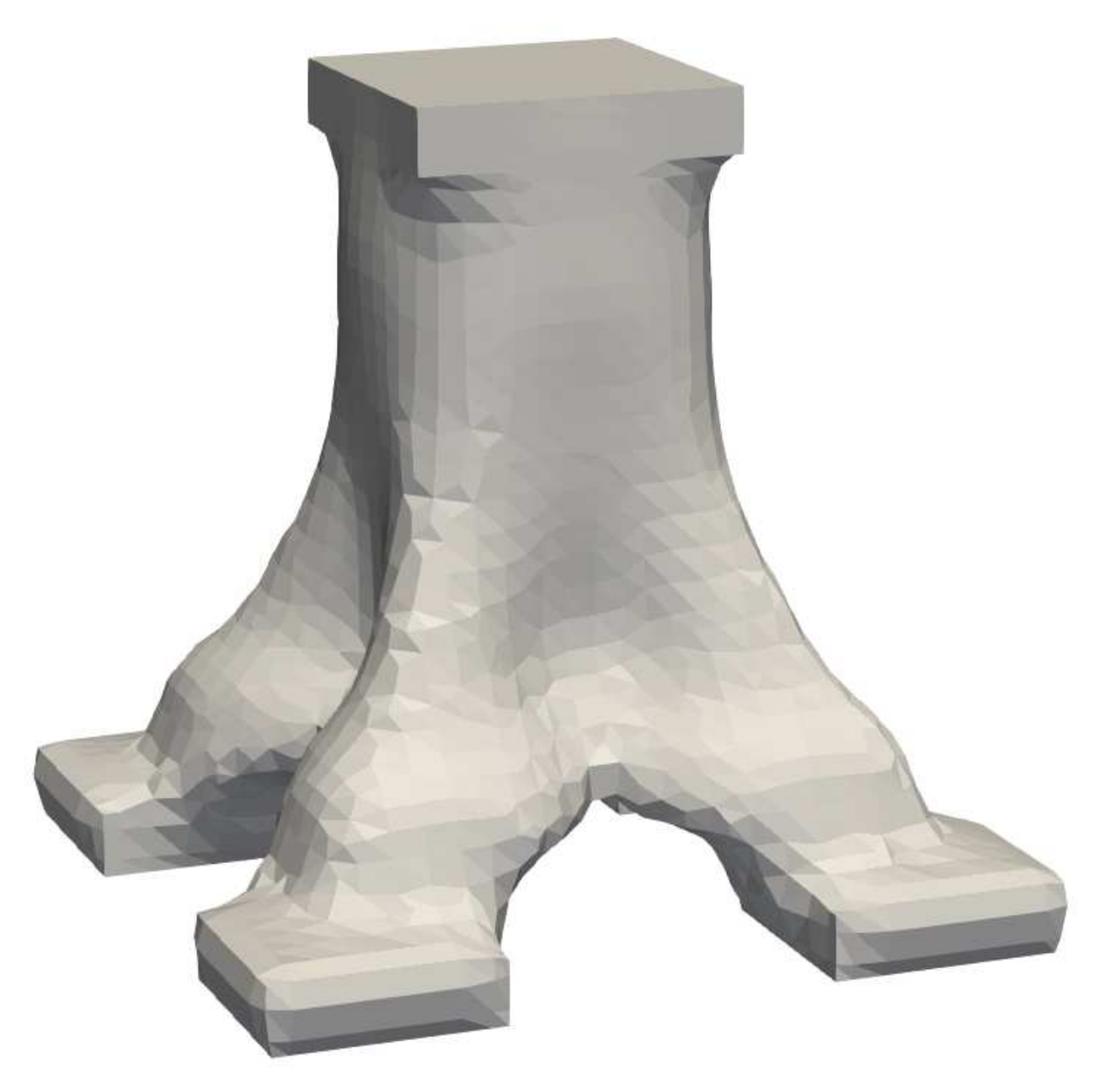} &
			\includegraphics[width=40mm]{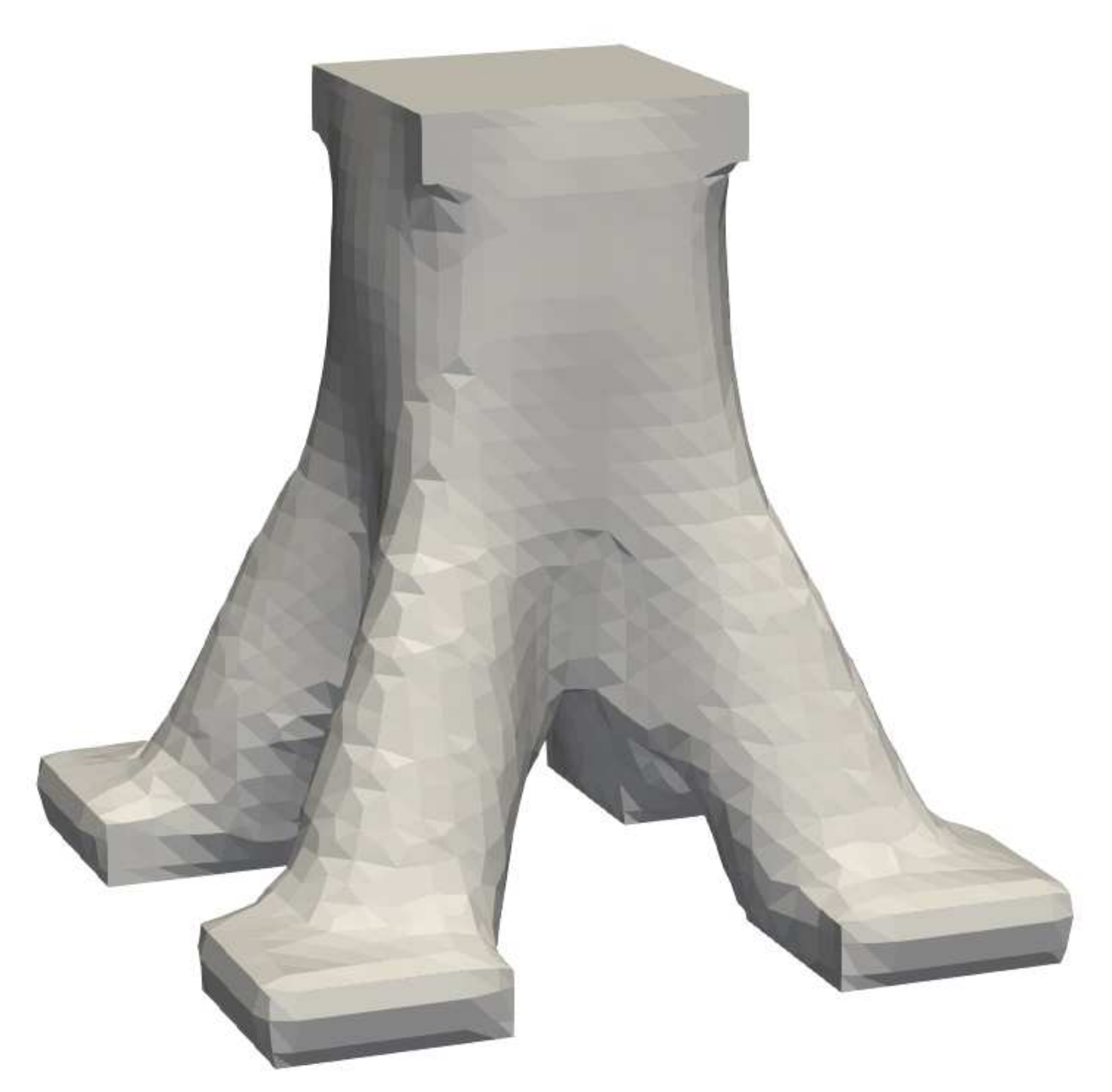} & \includegraphics[width=40mm]{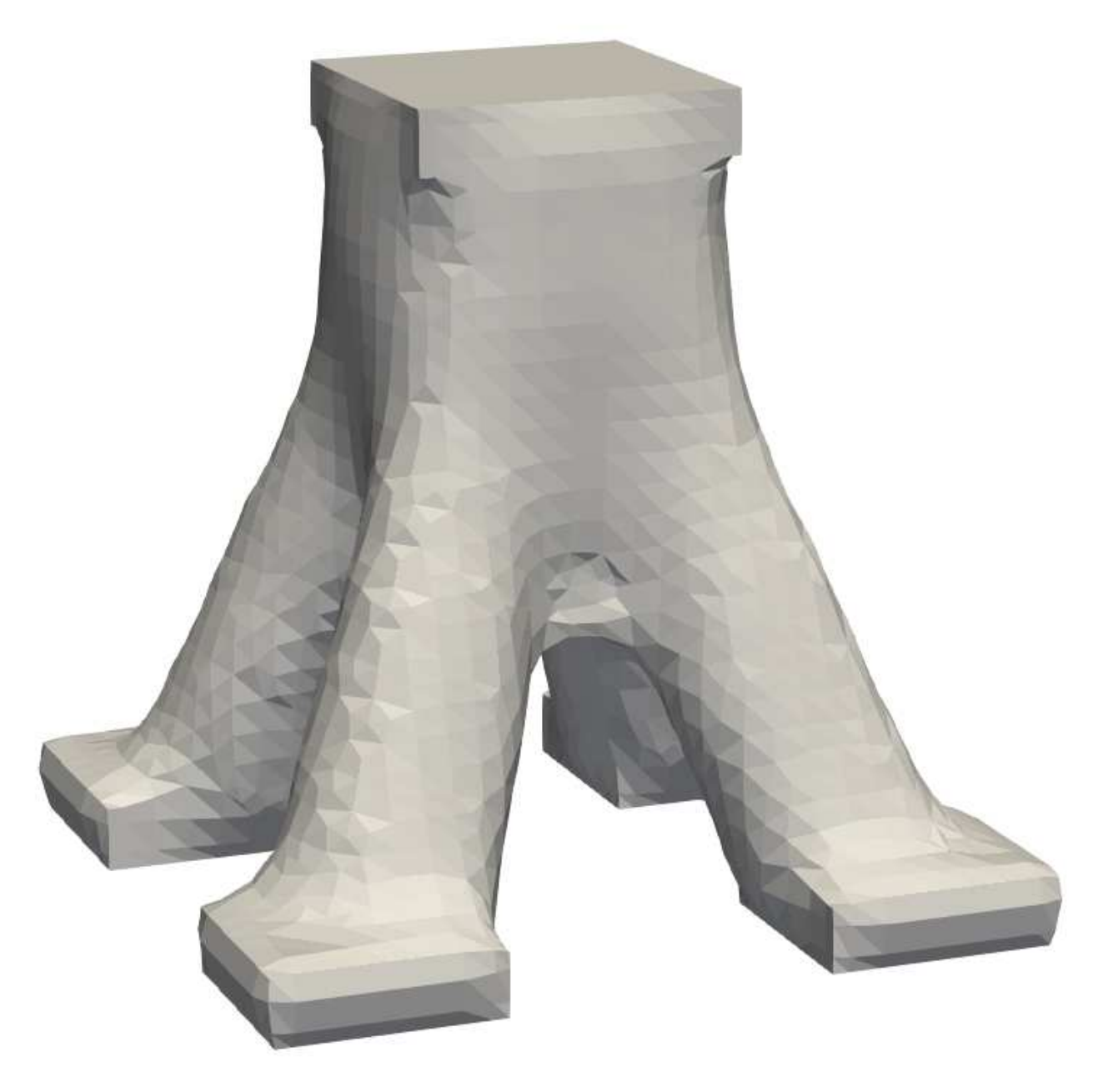} \\ 
			a) $\bitlength=2$, $\bitradius=3$, $\headradiusA=15$ & b) $\bitlength=5$, $\bitradius=3$, $\headradiusA=15$ & c) $\bitlength=10$, $\bitradius=3$, $\headradiusA=15$ \\   
			\includegraphics[width=40mm]{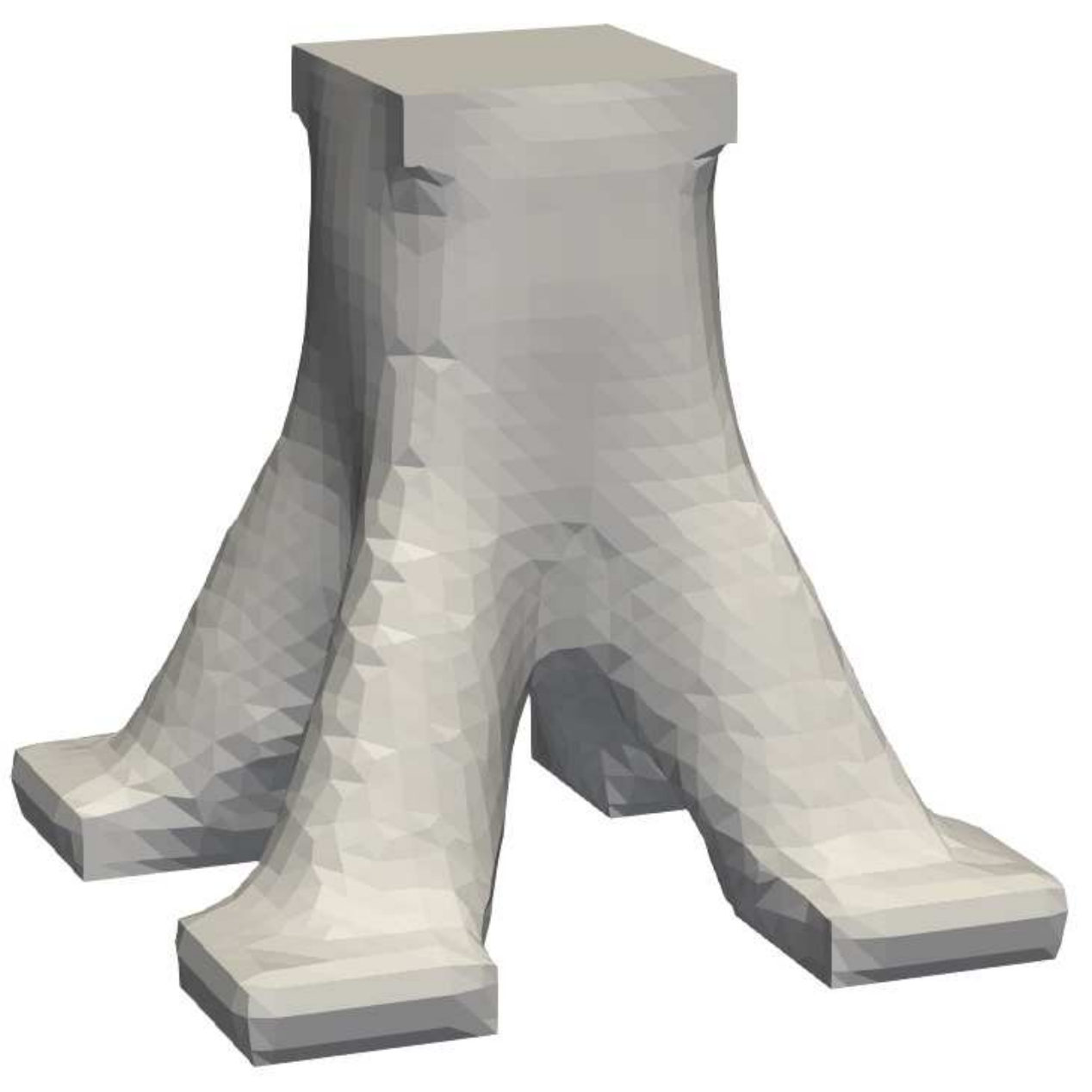} &
			\includegraphics[width=40mm]{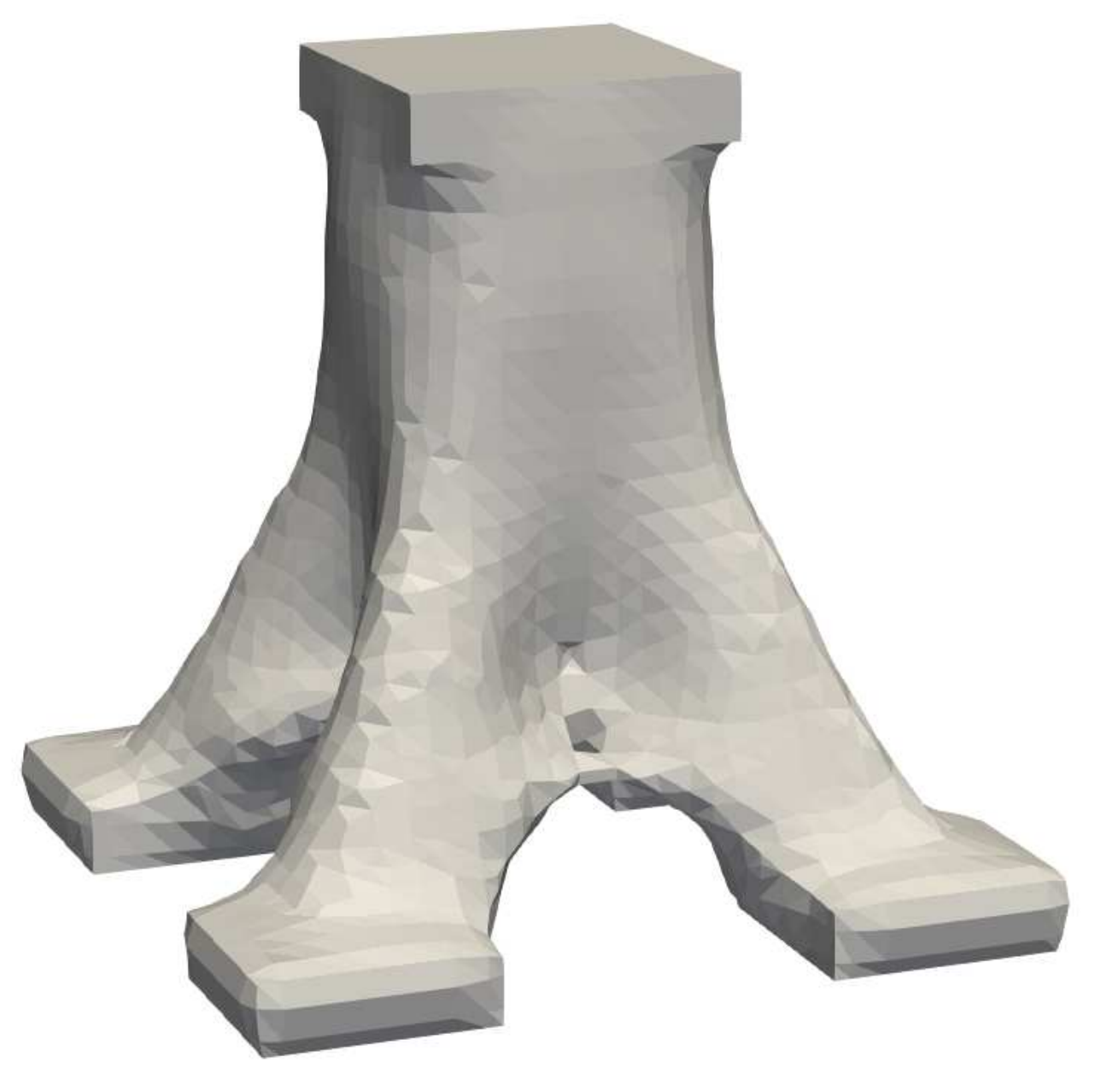} & \includegraphics[width=40mm]{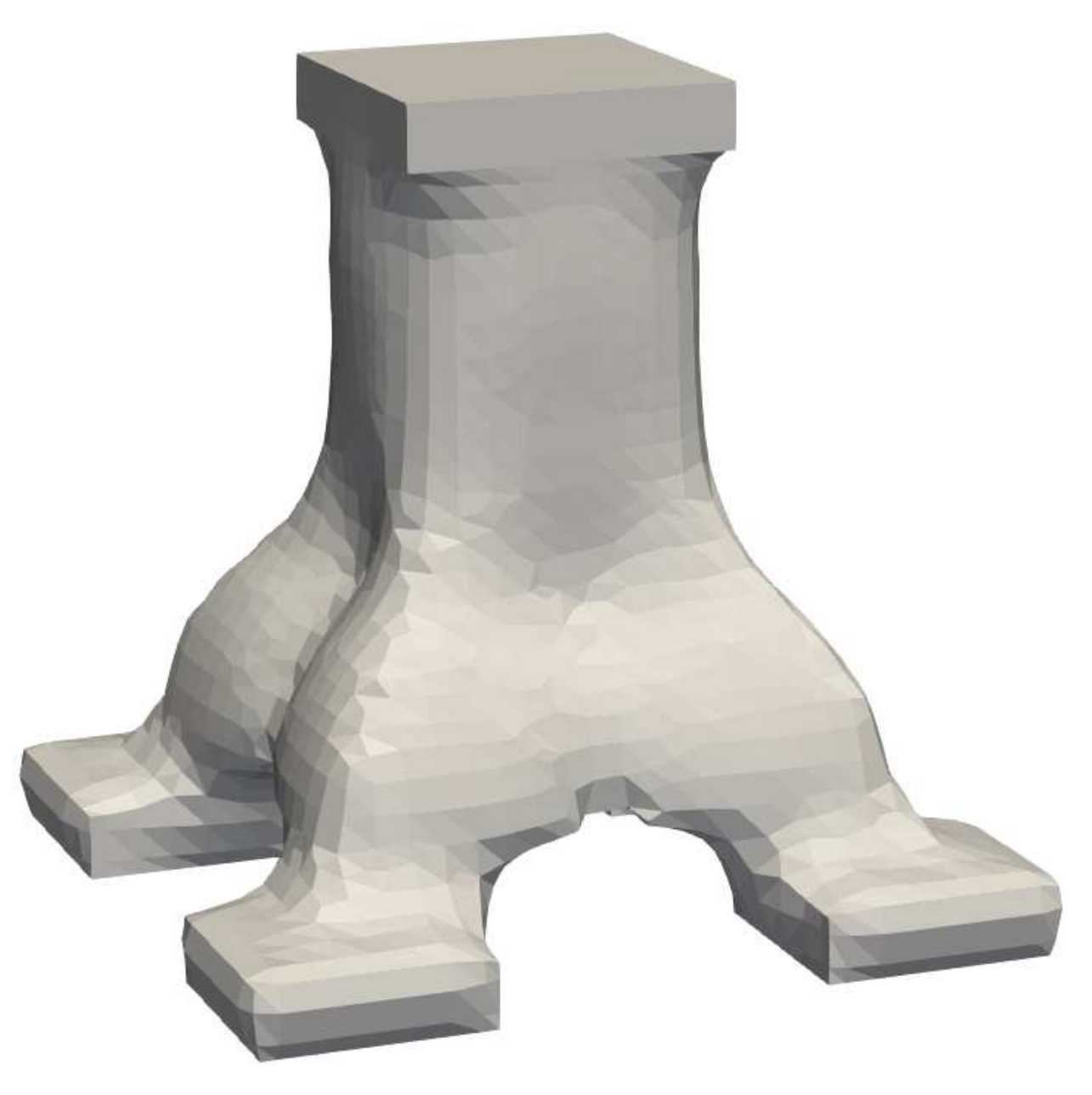} \\ 
			d) $\bitlength=3$, $\bitradius=3$, $\headradiusA=10$ & e) $\bitlength=3$, $\bitradius=3$, $\headradiusA=15$ & f) $\bitlength=3$, $\bitradius=3$, $\headradiusA=25$ \\
			\includegraphics[width=40mm]{figures/4sup3_b10h15r3-eps-converted-to.pdf} &
			\includegraphics[width=40mm]{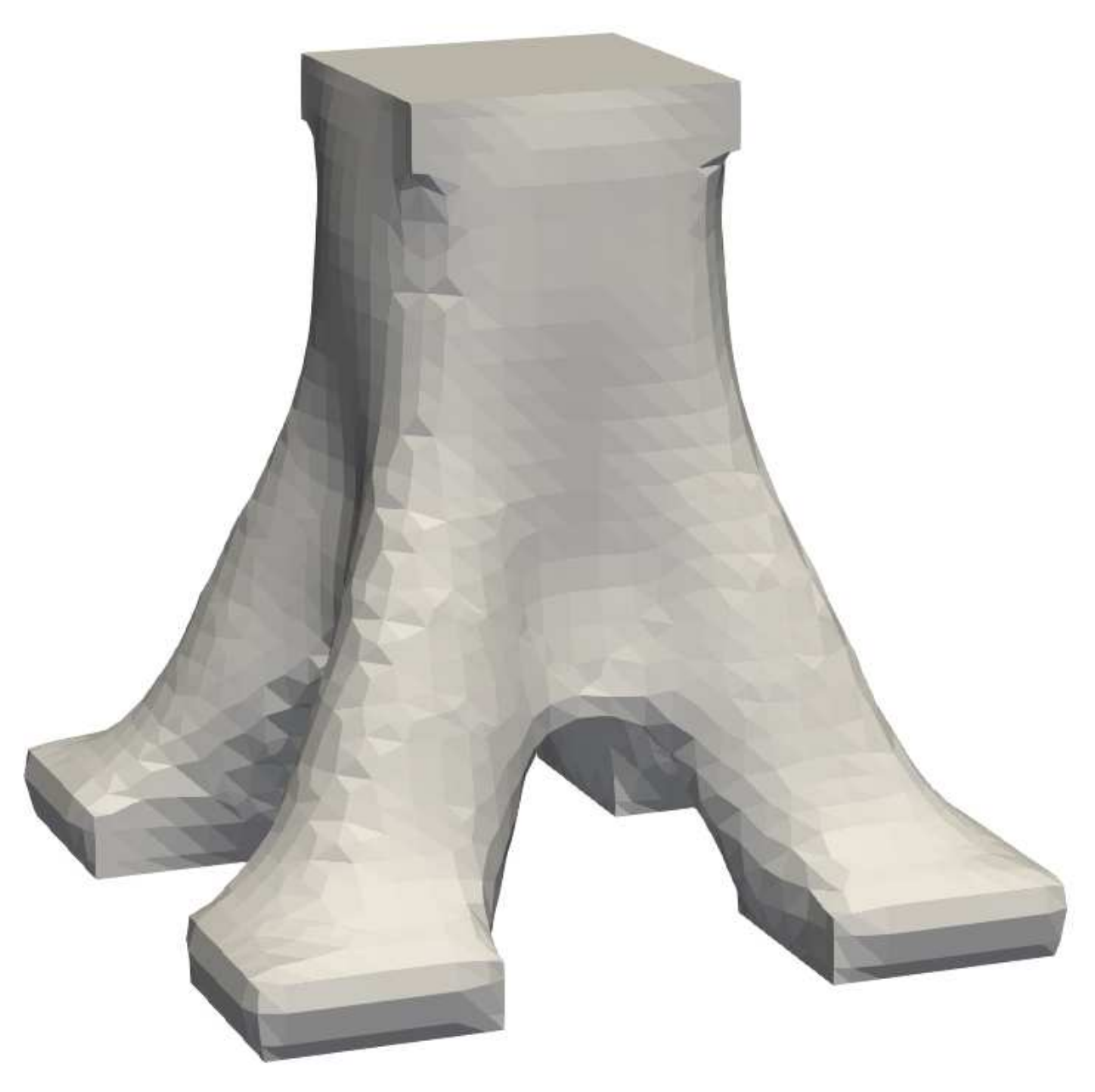} & \includegraphics[width=40mm]{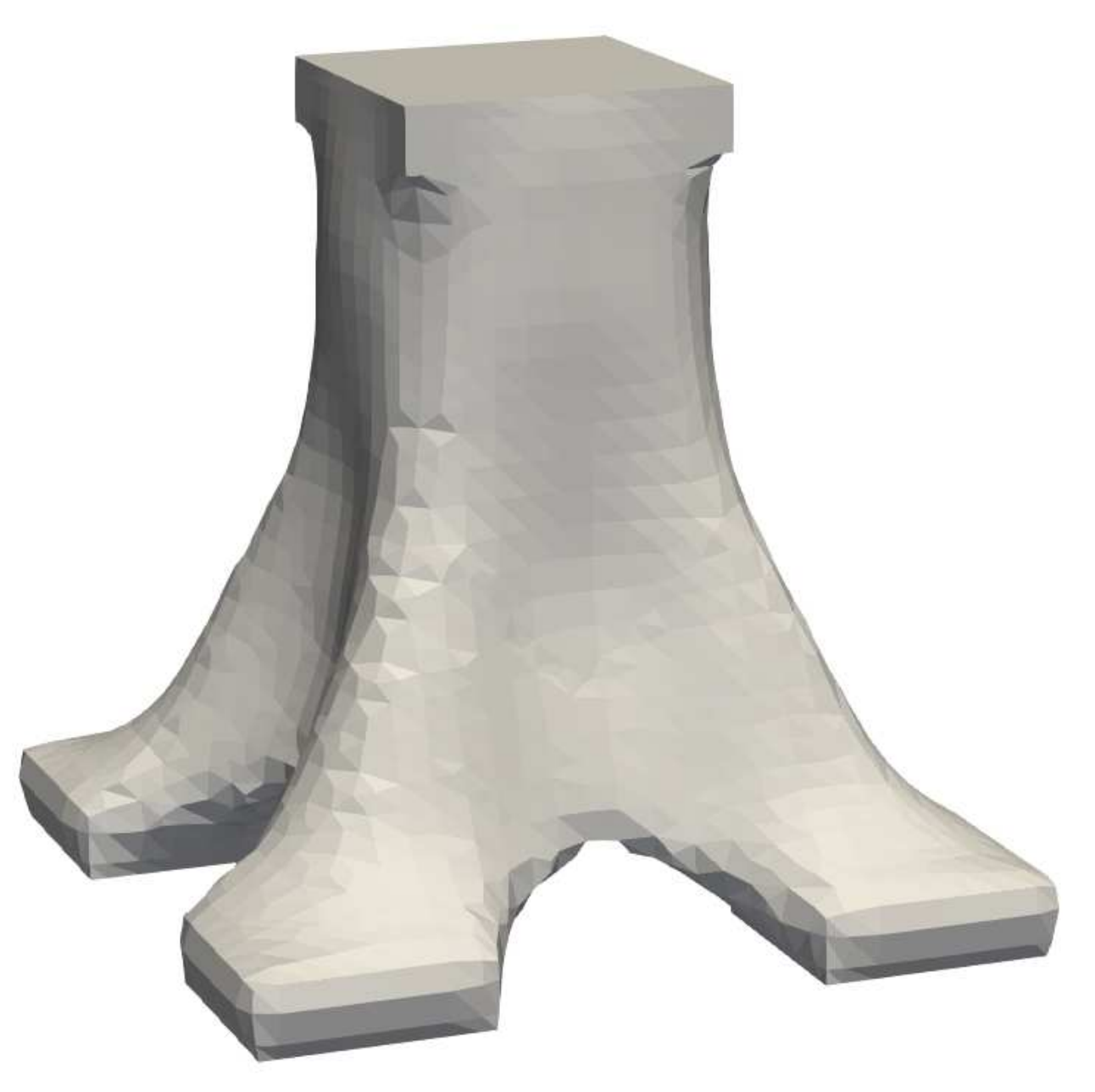} \\ 
			g) $\bitlength=10$, $\bitradius=3$, $\headradiusA=15$ & h) $\bitlength=10$, $\bitradius=7$, $\headradiusA=15$ & i) $\bitlength=10$, $\bitradius=10$, $\headradiusA=15$ \\  
		\end{tabular}
		\caption{Optimized parts with 3-axis milling using $\millingDirSet$: (+X,-X,+Y,-Y,+Z,-Z) and varying tool and head parameters.}
		\label{fig:3axisParams}
	\end{center}
\end{figure}

We ran experiments to test the effect of the various milling tool parameters on the {\bf SupportStruct} problem. Figure \ref{fig:3axisParams} (a-c) shows how the output varies as we modify the bit length. The access to the bottom cavity becomes reduced with shorter bits resulting in a squatter result (a). Figure \ref{fig:3axisParams} (d-f) shows how changing the head radius affects the result. In f) the effect is most noticeable where a large head prevents access to the geometry above the fixed plates and again squat result is obtained. In Figure \ref{fig:3axisParams} (g-i) the tool radius is varied and so the curvature of the resulting geometry is reduced as the tool increases in radius, giving a smoother result with fewer divots. 

\begin{figure}[h]
	\begin{center}
		\small
		\includegraphics[width=40mm]{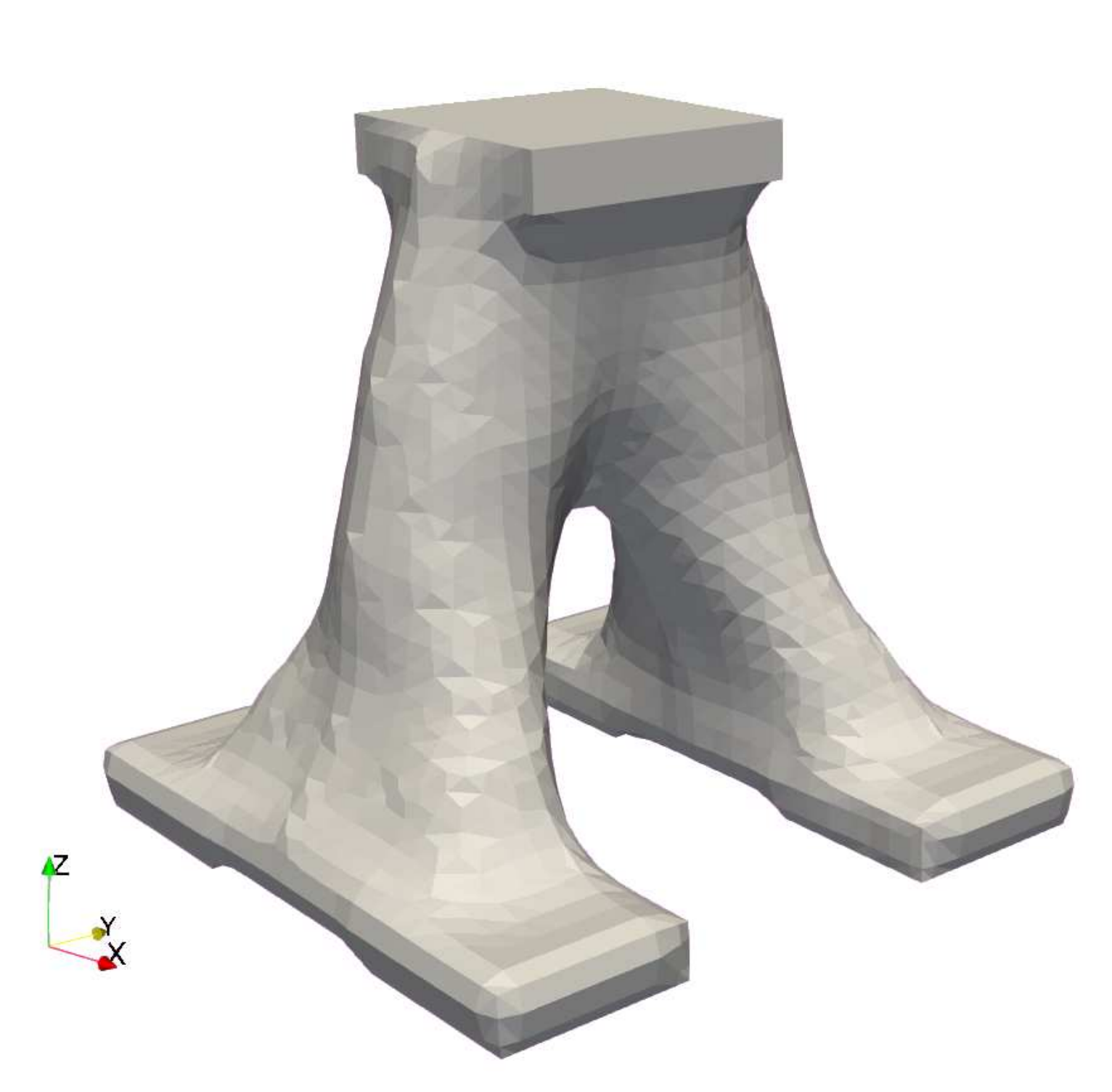}
		\includegraphics[width=40mm]{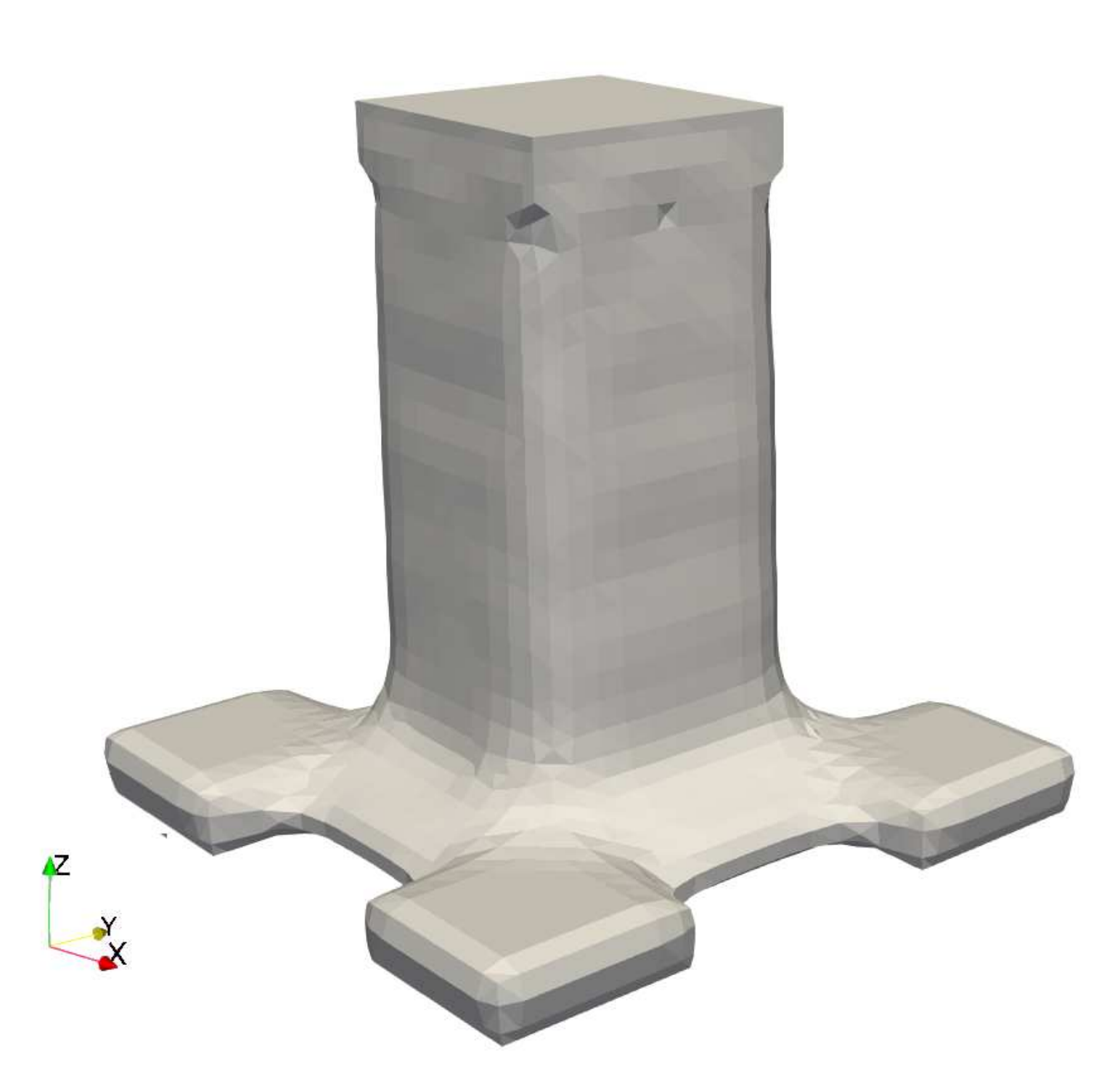}
		\caption{Optimized parts with 3-axis milling with varying milling directions. {\bf Left:} $\bitlength=20$, $\bitradius=3$, $\headradiusA=15$, $\millingDirSet$: (+X,-X). {\bf Right:} $\bitlength=50$, $\bitradius=3$, $\headradiusA=15$, $\millingDirSet$: (-Z)}
		\label{fig:3axisDirs}
	\end{center}
\end{figure}

We tested some variations of milling directions for 3-axis setups as shown in Figure \ref{fig:3axisDirs}. These illustrate how the preserved regions prevent removal of material that is inaccessible from the specified milling directions. We found that in general the milling constraint did not hinder the volume constraint satisfaction as long as the parameters of the bit allowed for enough material to be removed. For instance, the example in Figure \ref{fig:3axisDirs} right could not remove enough material to meet the volume target unless we specified a long enough bit length.

\subsection{Algorithm experiments}

\begin{figure}[h]
	\begin{center}
		\small
		\includegraphics[height=40mm]{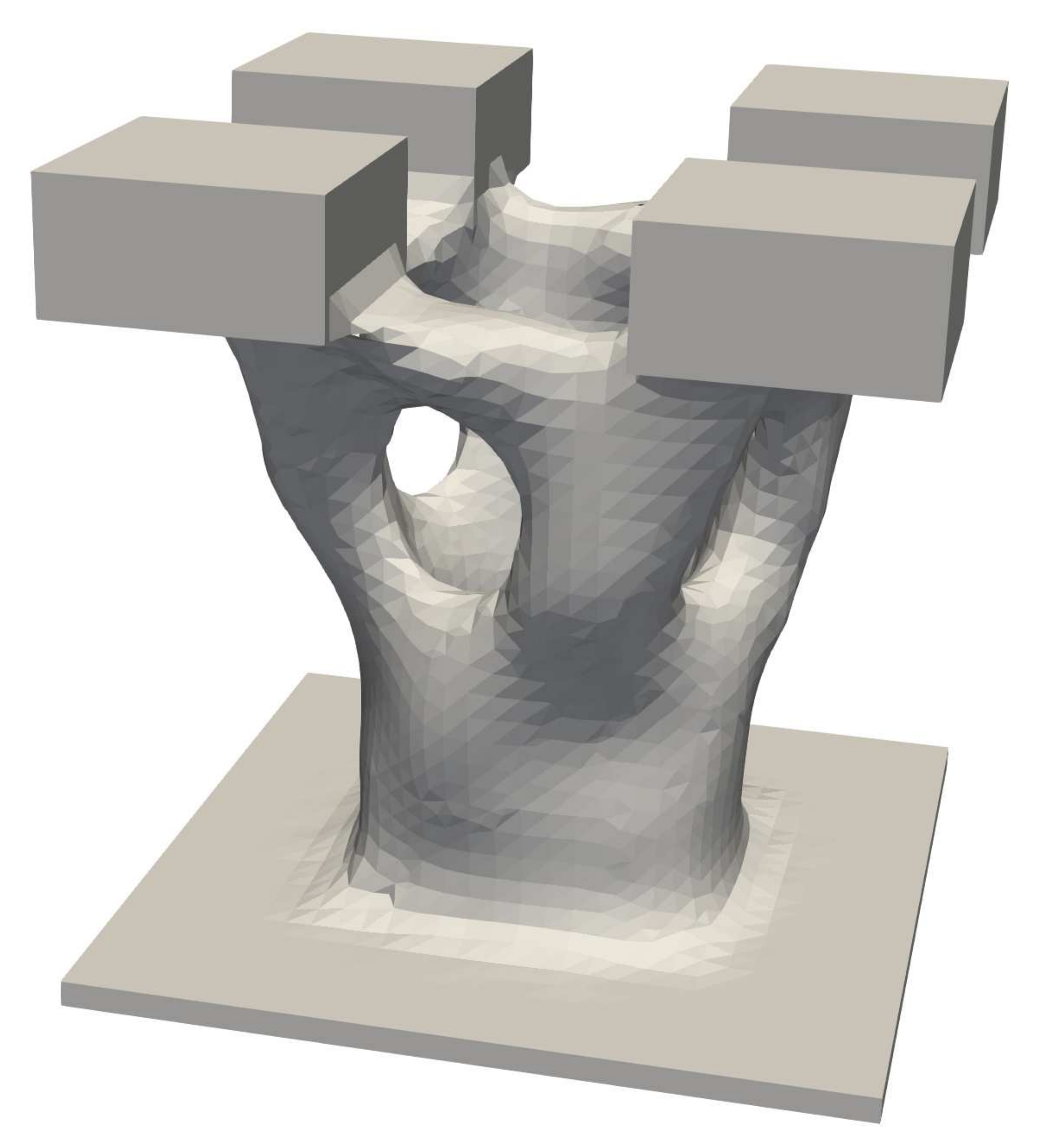}
		\includegraphics[height=40mm]{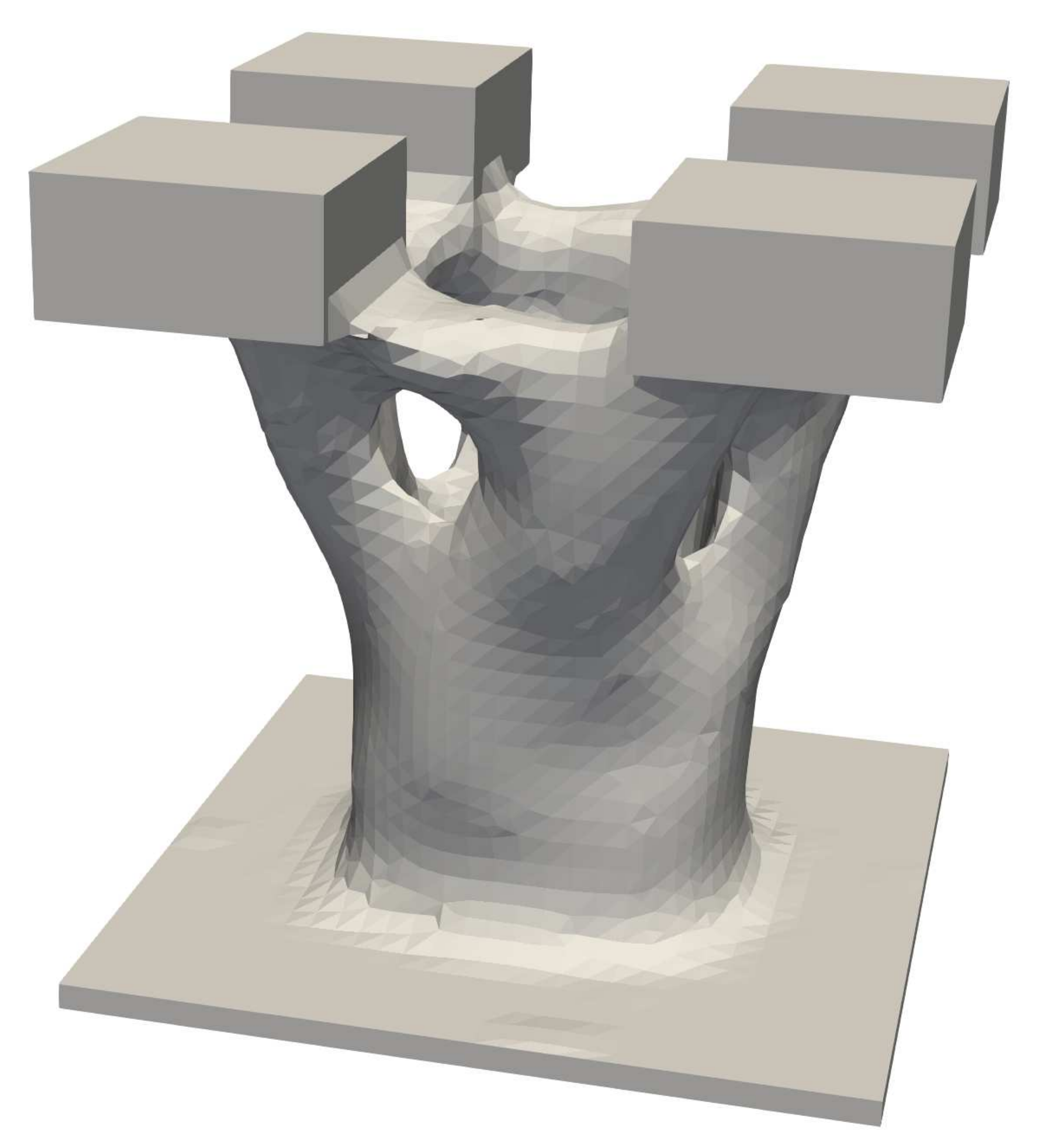}
		\includegraphics[height=40mm]{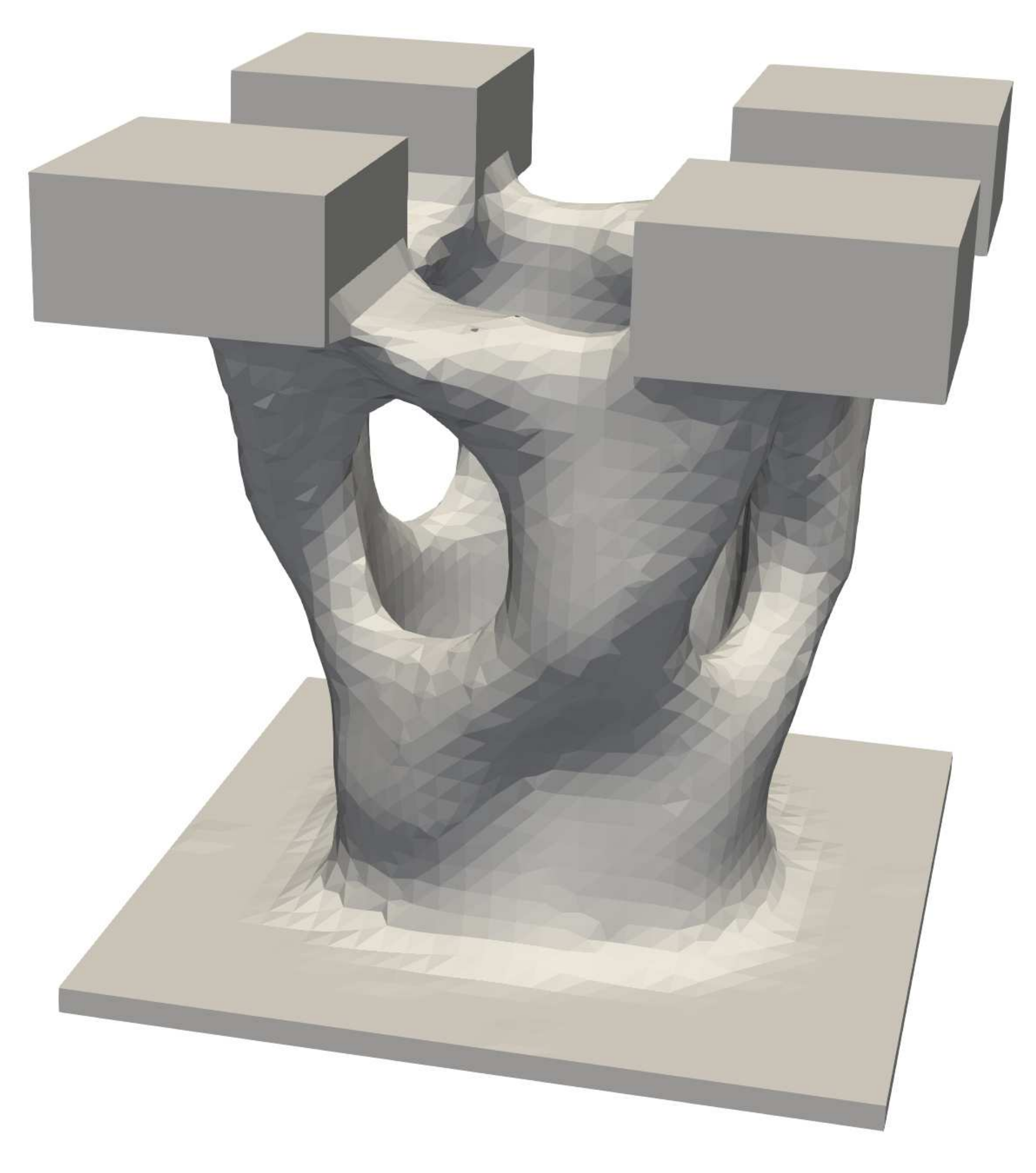}
		\includegraphics[height=40mm]{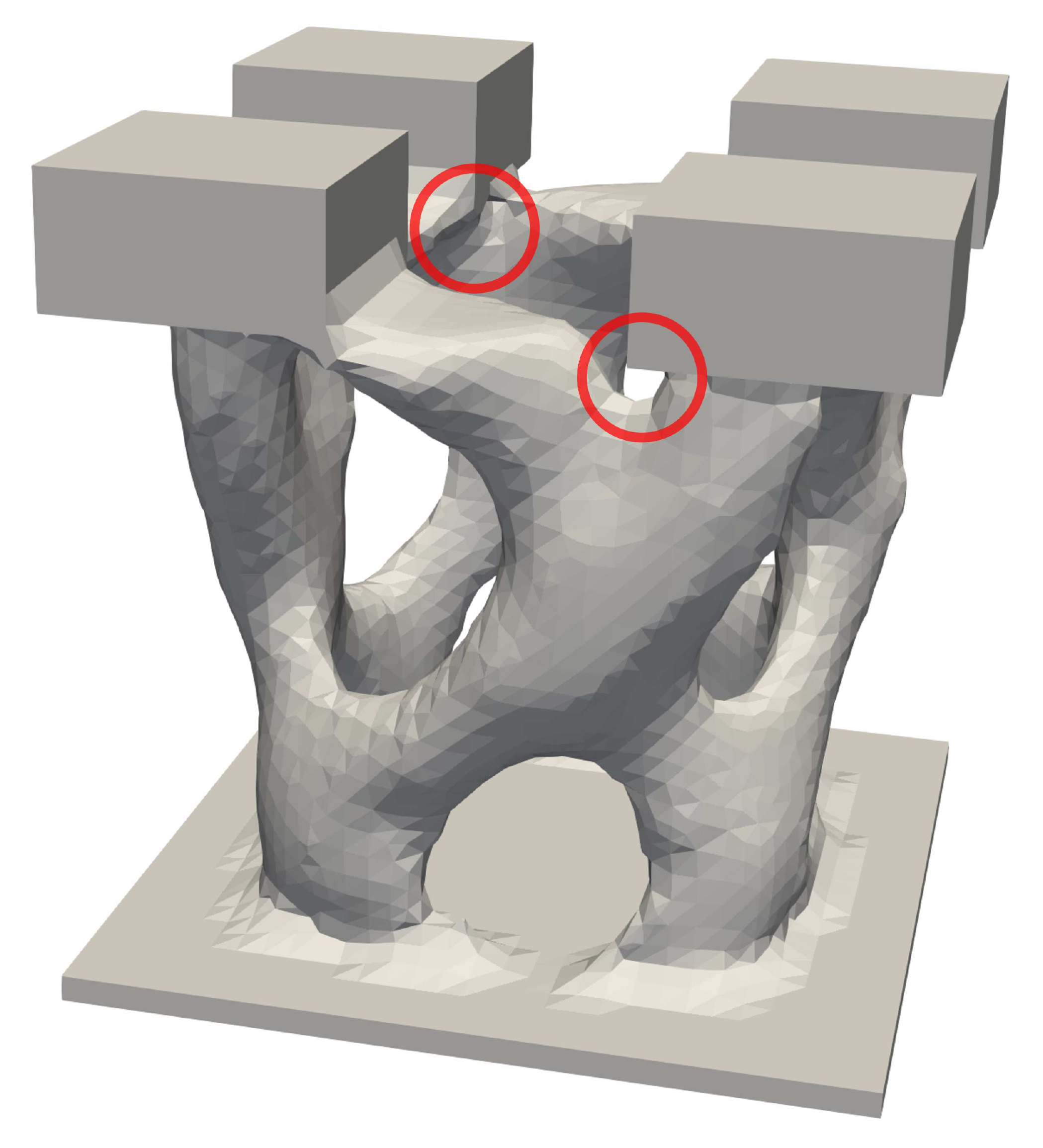}
		\caption{Search algorithm comparison. From left to right, Hemisphere search, normal search, heat search and no subtractive constraint. In the final image there are two areas highlighted where the output would not be manufacturable due to inaccessibility caused by overhangs and features that are too small for the bit to mill. }
		\label{fig:torqueResults}
	\end{center}
\end{figure}

We evaluated the performance of the 5-axis search algorithms on the {\bf TorqueStruct} example since the design space has a deep cavity, it serves as an illustration of the different $\filter$ search algorithms (see Table \ref{tab:table1}). Figure \ref{fig:torqueResults} shows the output after 40 iterations for each of the 5-axis search algorithms in section \ref{sec:impl} as well as an unconstrained topology optimization result. Each method converged to the target volume fraction, however the relative compliance and the geometry varied. As expected, the unconstrained result had the lowest compliance, but of the constrained versions the `Heat search' was able to identify surface elements deep inside the cylindrical cavity as accessible, while the `Normal search' was only able to remove material from the top and gradually opened the central cavity. The `Hemisphere search' also was not able to widen the cavity and resulted in the highest compliance.

In terms of performance, the `Normal search' had least impact on computation time as it does not need the heat equation solve and the `Hemisphere' method samples all 26 directions for each point. 

\begin{table}[h!]
	\begin{center}
		\label{tab:table1}
		\begin{tabular}{l|l|l|l}
			\textbf{Algorithm} & \textbf{Relative compliance} & \textbf{Volume} & \textbf{Avg time(s)/iter}\\
			\hline
			Hemisphere & 1.000 & 0.302 & 8.641\\
			Normal & 0.968	& 0.301	& 8.333\\
			Heat & 0.901	& 0.301	& 14.179\\
			No subtractive & 0.659	& 0.302	& 7.359\\
		\end{tabular}
		\caption{Comparison of the three accessibility search algorithms as well as the baseline optimization unconstrained by any subtractive constraint.}
	\end{center}
\end{table}

\begin{figure}[h]
	\begin{center}
		\small
		\includegraphics[width=120mm]{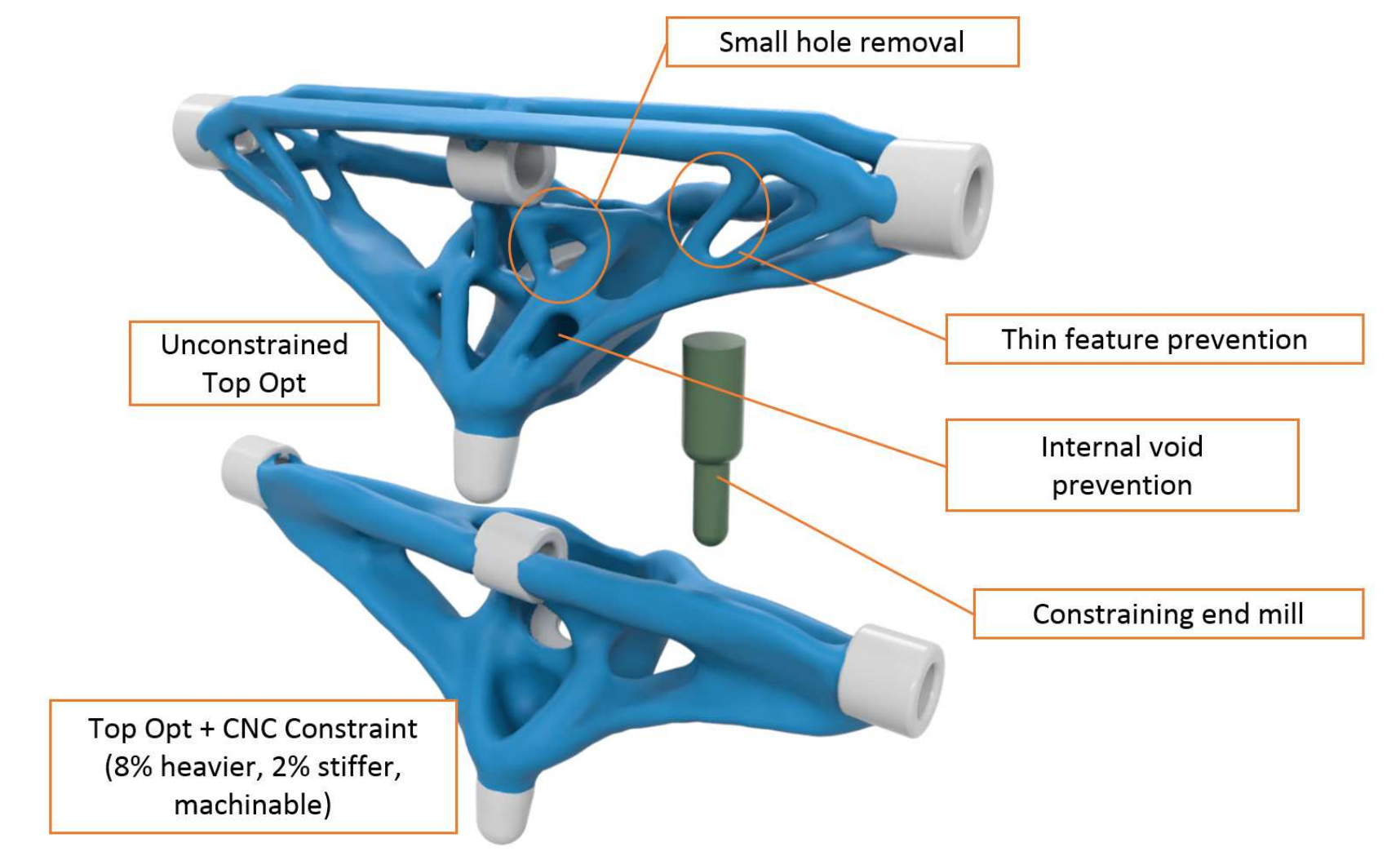}
		\includegraphics[width=100mm]{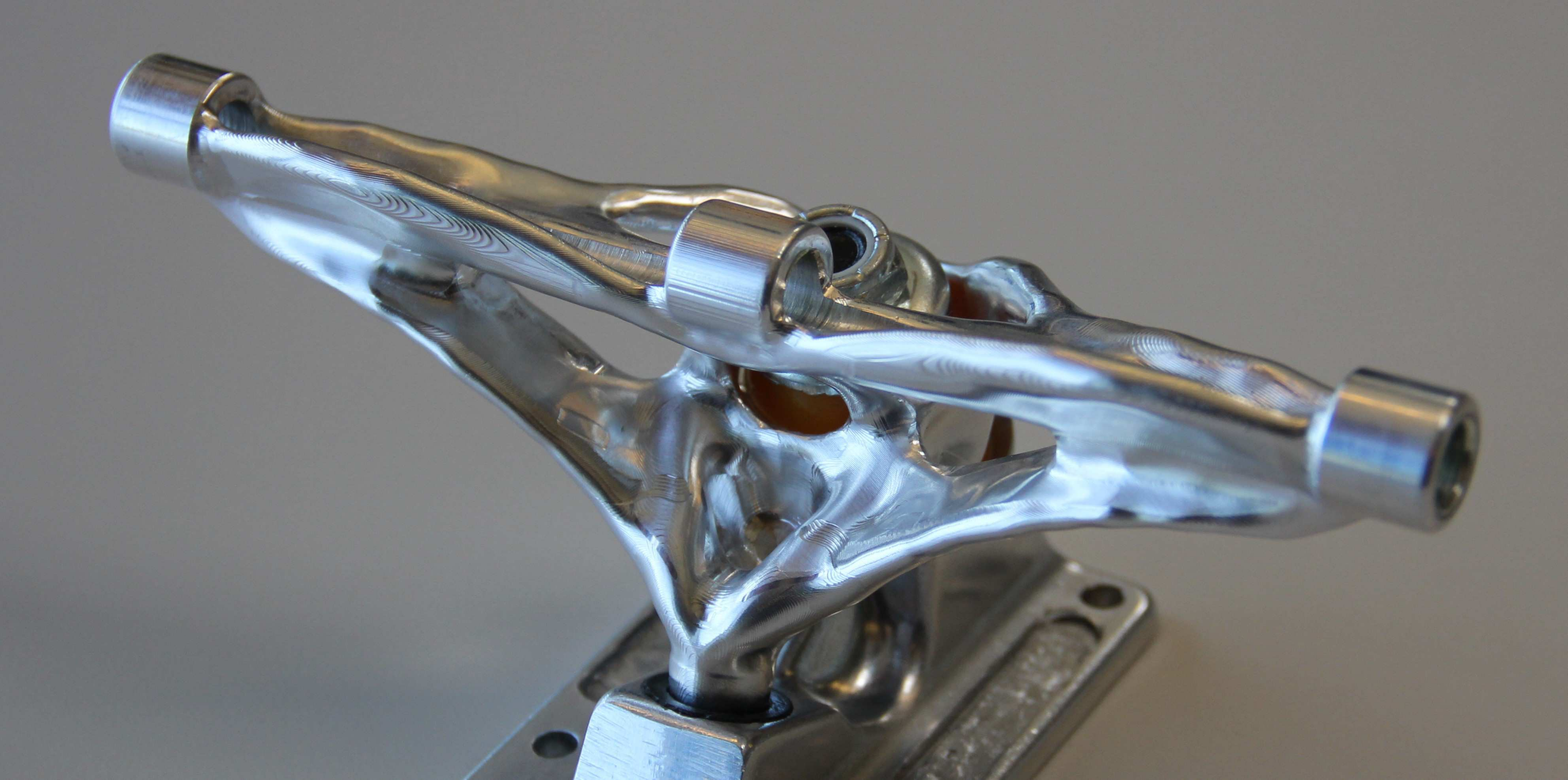}
		\caption{{\bf Top:} Comparison between unconstrained topology optimization and 5-axis CNC milling constrained topology optimization. {\bf Bottom:} Manufactured result using a 5-axis CNC milling procedure.}
		\label{fig:skate}
	\end{center}
\end{figure}

Figure \ref{fig:skate} shows a comparison between unconstrained topology optimization on a skateboard truck (above) and the result of topology optimization with the 5-axis CNC constraint described above (using `Heat Search'). Note that the unconstrained version has a number of features that would be impossible to mill, such as small holes, thin features and cage structures that would not be accessible by the tool. The constrained version avoids all of these areas and results in a much more `compressed' result with fewer struts and all the cavities are open to attack from the tool. The constrained result is just 8\% heavier and 2\% stiffer than the unconstrained version. To validate this result we manufactured the design using a 5-axis Matsuura MX-330 CNC milling machine. We used a $\frac{1}{4}$ inch bull nose end mill for roughing and a $\frac{1}{8}$ inch ball nose end mill for finishing. 

\begin{figure}[h!]
	\begin{center}
		\small
		\includegraphics[width=120mm]{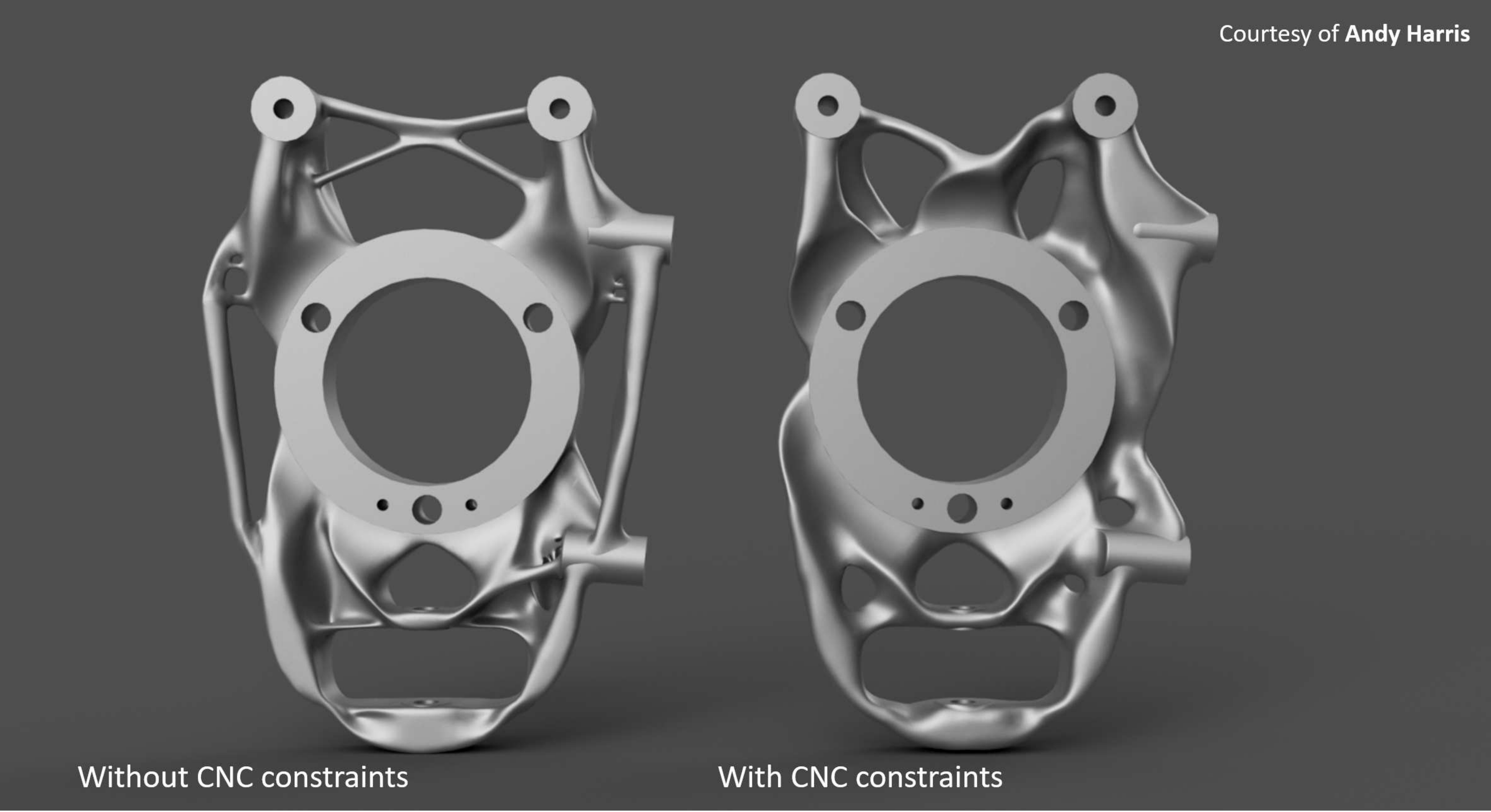}
		\includegraphics[width=120mm]{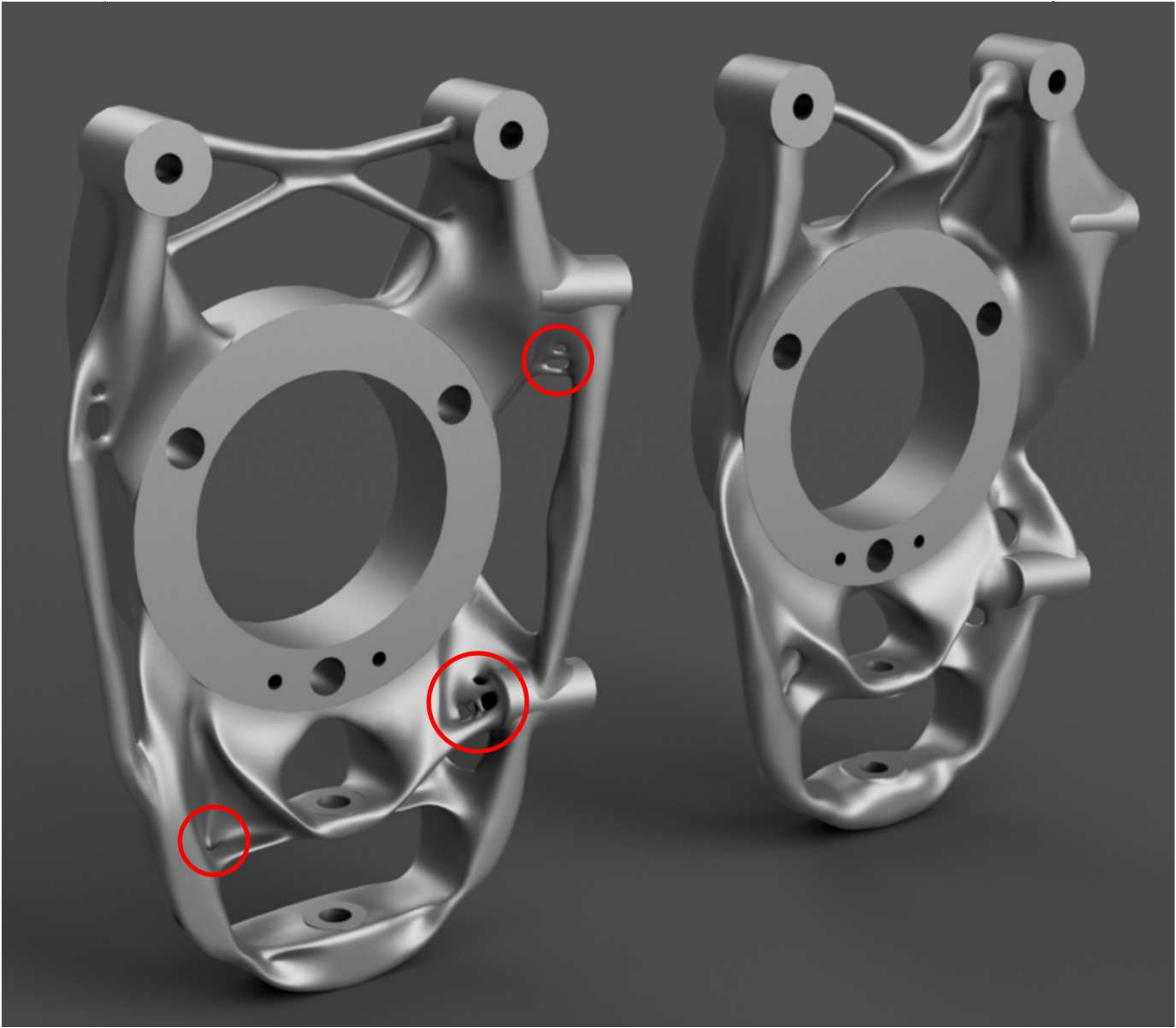}
		\includegraphics[width=85mm]{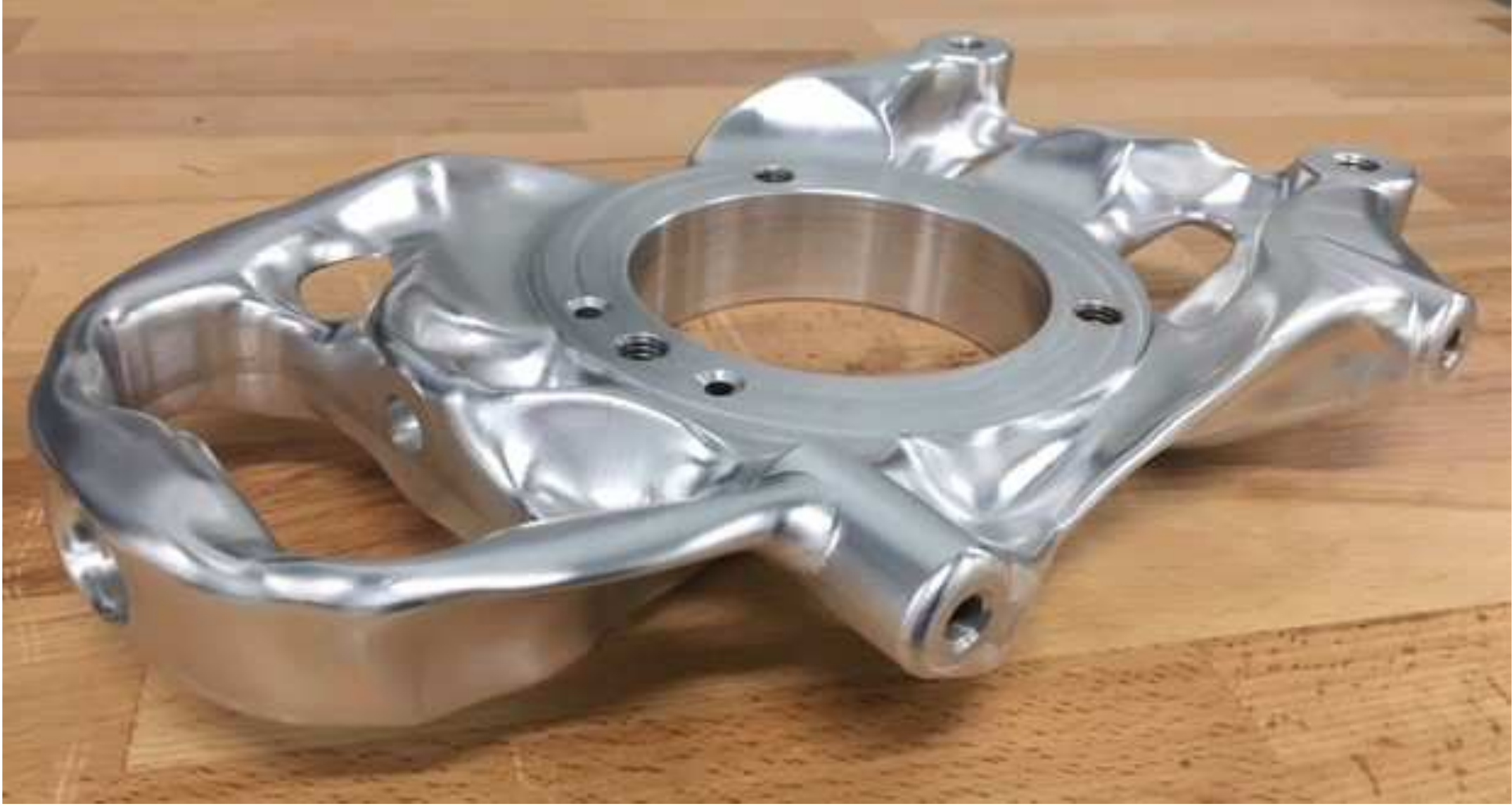}
		\includegraphics[width=34mm]{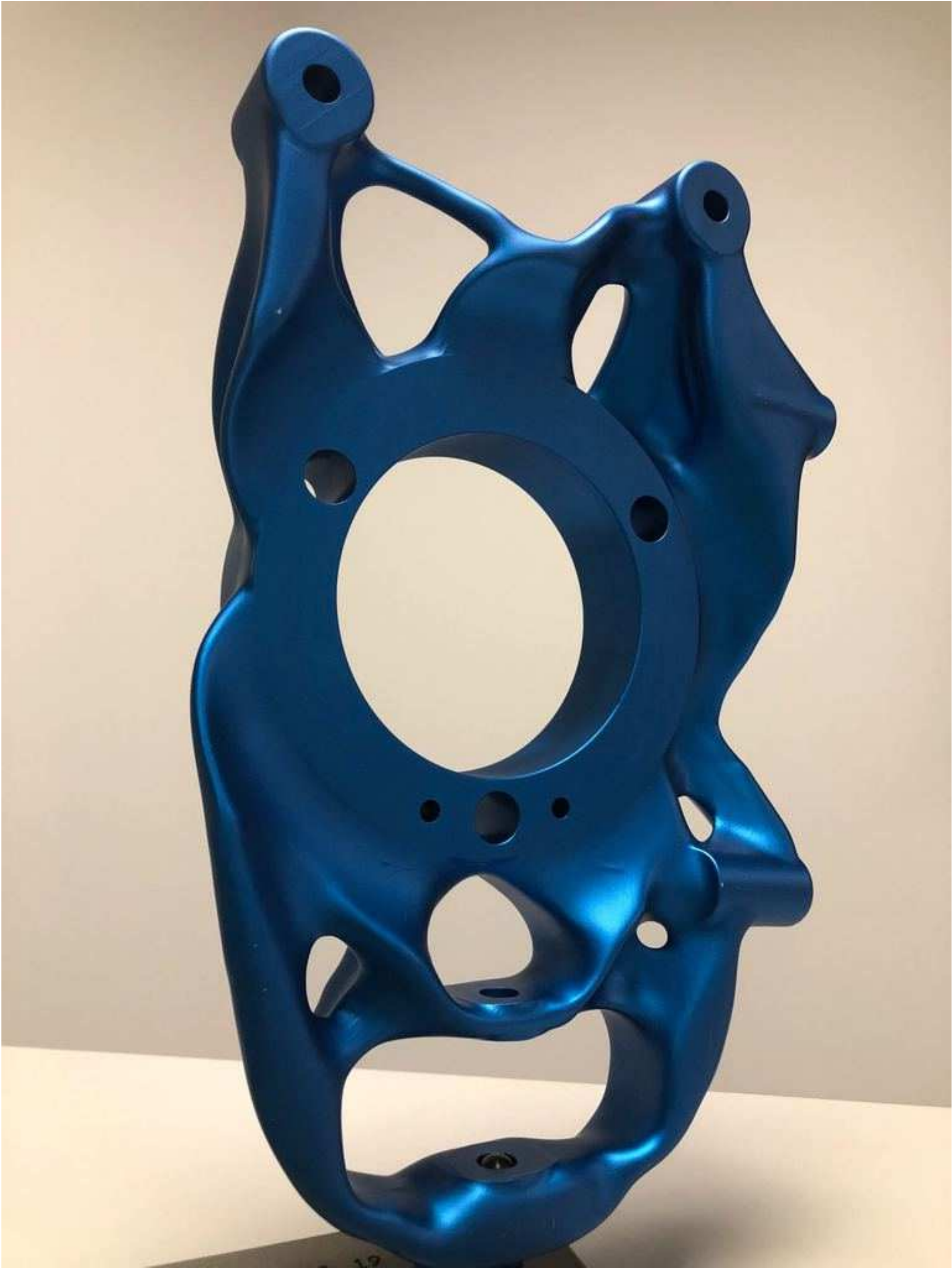}
		\caption{{\bf Left:} Unconstrained topology optimization of race car upright with highlighted non-machinable regions, {\bf Right:} 5-axis CNC milling constrained topology optimization. {\bf Bottom:} Manufactured result using a 5-axis CNC milling procedure.}
		\label{fig:BAC}
	\end{center}
\end{figure}

We demonstrated the algorithm with another case study for a racing car upright. Figure \ref{fig:BAC} shows the unconstrained topology optimization results along highlighted regions where the curvature is too high or the tool bit cannot access the surface. Alongside this is the constrained 5-axis result that again is more compact without inaccessible undercuts and larger features. We also validated the output by generating a fully machined version of the result shown at the bottom of Figure \ref{fig:BAC}.

\section{Conclusion}

We have demonstrated a novel method for topology optimization with a manufacturing constraint suitable for subtractive milling machines. Our constraint allows the user to define a milling strategy for 3-axis or 5-axis machines as well as the shape of the tool bits and machine heads that will be used to manufacture the part. We have shown results for variations of the constraint parameters.

\subsection{Extensions}

Here we explore several potential extensions to the method as well as future work.

It is possible to consider an even wider range of tool bits and heads by adding additional head layers (approximating the head with two or more capped cylinders), or additionally by modeling a tapered frustum instead of a cylinder. Frustum intersection can be achieved by modifying the ray casting function. Another extension would be to allow the user to specify a set of potential tools at their disposal and then the optimization can do accessibility testing for each tool and choose the tool with highest $\filter$ value.

One of the limitations of the strict algorithm is that the velocity function is clamped such that no growth can occur in the part. Often a topology optimization will require the shifting a subset of the part region or the growth of a subset of the part while another region shrinks. This limits the effectiveness of the topology optimization. We presented the relaxed algorithm as one method to alleviate this restriction. Another potential method is to detect regions with positive velocity and then to backtrack to the previous iteration and set $\filter = 0$ in that region so no volume is removed, thus anticipating the growth. 

The algorithms presented here enforce accessibility for the specified machine parameters, this should ensure that the part is millable however it may be very slow to manufacture. Optimizing for milling time is a very interesting extension for future work.

\section{Conflict of interest statement}

The authors are employed at Autodesk Inc.

\section{Replication of Results}

The results were obtained using optimization algorithms defined in section \ref{sec:optAlg} using the level set framework from \cite{Museth13} and the pseudo code algorithms are provided in the appendix for ease of replication.

\section{Appendix}

\small
\begin{algorithm}[H]
	\KwIn{Level set, rayStart, rayDirection, levelSetValue, hitPoint}
	\KwResult{Returns true if the ray crosses the level set at levelSetValue and stores the intersection point in hitPoint}
	\caption{LevelSetRayCast}
	\label{alg:rayCast}
\end{algorithm}

\begin{algorithm}[H]
	\KwIn{Level set $\partDomain$, offset value $\offset$}
	\KwResult{Applies an offset of $\offset$ to the signed distance function for $\partDomain$ such that the zero-levelset is uniformly advanced outwardly in the direction of  $\offset\normal$, where $\normal$ is the surface normal. If the level set is narrow band then the narrow band must be updated such that it centers on the new zero-level set.}
	\caption{LevelSetOffset}
	\label{alg:levelSetOffset}
\end{algorithm}

\begin{algorithm}[H]
	\KwIn{Level set $\partDomain$, scalar field $\speed$, time $\advectTime$}
	\KwResult{Solves the Hamilton-Jacobi equation on the signed distance function with the speed defined by the scalar field $\speed$ over a duration $\advectTime$. Note that a velocity extension method improves the stability of advection \cite{Adal99}.}
	\caption{LevelSetAdvection}
	\label{alg:levelSetAdvection}
\end{algorithm}

\begin{algorithm}[H]
\KwIn{Level set $\partDomain$, close value $\offset$}
	\KwResult{Morphological operation that performs an offset by $\offset$ followed by an inset or negative offset of $\offset$. The result is a removal of holes of radius $\offset$.}
	\caption{LevelSetClose}
	\label{alg:levelSetClose}
\end{algorithm}

\begin{algorithm}[H]
	\KwIn{Level set $\partDomain$, surface normal $\normal(\surfacePt)$, offset point $\toolTip$, milling direction $\millingDir_i$, filter value $\filter(\surfacePt)$}
	\KwResult{Returns accessibility boolean, the updated filter $\filter(\surfacePt)$ for this $\millingDir_i$ and the intersection point $\intPt$}
	\If{$\normal(\surfacePt) \cdot \millingDir_i < 0$}{
		hit, $\hitPt$ = LevelSetRayCast($\partDomain$,$\toolTip$,$-\millingDir_i$,$\bitradius$,$\intPt$)\;
		\uIf{!hit}{
			hit, $\hitPt$ = LevelSetRayCast($\partDomain$,$\toolTip-\millingDir_i(\bitlength+\headradiusA)$,$-\millingDir_i$,$\headradiusA$,$\intPt$)\;
			\If{!hit}{
				$\filter(\surfacePt) = \max{(\filter(\surfacePt),-\millingDir_i \cdot \normal(\surfacePt)}$)\;
				\Return (accessible = true, $\filter(\surfacePt), \intPt$)\;
			}
		}
		\Else
		{
			\Return (accessible = false, $\filter(\surfacePt), \intPt$)\;
		}
	}
	\Return (accessible = false, $\filter(\surfacePt), \intPt$)\;
	\caption{MillingTest: Level set evaluation for $\millVolume_i(\surfacePt) \cap \partDomain$}
	\label{alg:millTest}
\end{algorithm}

\bigskip

\begin{algorithm}[H]
	\KwIn{Level set of $\partDomain$, $\normal$, set of milling directions $\millingDirSet$}
	\KwResult{Evaluation of $\filter$}
	Extend narrow-band of $\partDomain$ up to $\headradiusA$\;
	\For{$\surfacePt$ on boundary of $\partDomain$}{
		$\toolTip = \surfacePt + \normal \cdot \bitradius$\;
		$\filter(\surfacePt) = 0$\;
		\For{$\millingDir_i$ in $\millingDirSet$}{
			accessible, $\filter(\surfacePt)$, $\intPt$ = MillingTest($\partDomain$,$\normal(\surfacePt)$,$\toolTip$,$\millingDir_i$,$\filter(\surfacePt)$)\;
		}
	}
	\Return $\filter$\;
	\caption{MillingConstraint for 3-axis}
	\label{alg:filter}
\end{algorithm}

\bigskip

\begin{algorithm}[H]
	\KwIn{Level set of $\partDomain$, $\normal$}
	\KwResult{Evaluation of $\filter$}
	Extend narrow-band of $\partDomain$ up to $\headradiusA$\;
	\For{$\surfacePt$ on boundary of $\partDomain$}{
		$\toolTip = \surfacePt + \normal \cdot \bitradius$\;
		accessible = false\;
		$\filter(\surfacePt) = 0$\;
		$\millingDir = -\normal$\;
		\While{!accessible}{
			accessible, $\filter(\surfacePt)$, $\intPt$ = MillingTest($\partDomain$,$\normal(\surfacePt)$,$\toolTip$,$\millingDir$,$\filter(\surfacePt)$)\;
			$\intPt' = \intPt$\;
			\Do{$\sdf(\intPt') - \sdf(\intPt) > 0$}{
				$\intPt' = \intPt' + \normal(\intPt)$\;
			}
			$\millingDir = \frac{\intPt' - \toolTip}{||\intPt'-\toolTip||}$\;
		}
	}
	\Return $\filter$\;
	\caption{MillingConstraint for 5-axis with NormalSearch}
	\label{alg:normalSearch}
\end{algorithm}

\bigskip

\begin{algorithm}[H]
	\KwIn{Level set of $\partDomain$, $\normal$}
	\KwResult{Evaluation of $\filter$}
	Extend narrow-band of $\partDomain$ up to $\headradiusA$\;
	$\TempField$ = SolveHeatEquation($\partDomain^+$)\;
	\For{$\surfacePt$ on boundary of $\partDomain$}{
		$\toolTip = \surfacePt + \normal \cdot \bitradius$\;
		$\heatPt = \toolTip$\;
		accessible = false\;
		$\filter(\surfacePt) = 0$\;
		$\millingDir = -\normal$\;
		\While{!accessible}{
			accessible, $\filter(\surfacePt)$, $\intPt$ = MillingTest($\partDomain$,$\normal(\surfacePt)$,$\toolTip$,$\millingDir$,$\filter(\surfacePt)$)\;
			$\heatPt = \heatPt + \heatEps \frac{\TempFieldGrad(\heatPt)}{||\TempFieldGrad(\heatPt)||}$\;
			$\millingDir = \frac{\TempFieldGrad(\heatPt)}{||\TempFieldGrad(\heatPt)||}$\;
		}
	}
	\Return $\filter$\;
	\caption{MillingConstraint for 5-axis with HeatSearch}
	\label{alg:heatSearch}
\end{algorithm}

\bigskip

\begin{algorithm}[H]
	\KwIn{Initial part domain $\partDomain$, Boundary conditions, Objective function, Milling parameters $\bitradius$,$\bitlength$ and $\headradiusA$, set of milling directions $\millingDirSet$}
	\KwResult{Optimized version of $\partDomain$}
	\While{!converged}
	{
		EvaluateSensitivities($\partDomain$,Boundary conditions,Objective function)\;
		$\speed$ = ComputeShapeDerivative()\;
		$\eta$ = MillingConstraint($\partDomain$,$\normal$,$\millingDirSet$,maxIters)\;
		\For{$\surfacePt$ on boundary of $\partDomain$}{
			\If{$\speed(\surfacePt) > 0$}{$\eta(\surfacePt) = 0$}
		}
		$\eps = $ LineSearch($\partDomain$,$\speed$)\;
		LevelSetAdvection($\partDomain$,$\eta\speed$,$\eps$)\;
	}
	LevelSetClose($\partDomain$,$\bitradius$)\;
	\caption{Inner Loop of the Augmented Lagrangian Algorithm for Level Set Topology Optimization --- CNC Constraint Strict Algorithm}
	\label{alg:strictAlgorithm}
\end{algorithm}

\bigskip

\begin{algorithm}[H]
	\KwIn{Initial part domain $\partDomain$, Boundary conditions, Objective function, Milling parameters $\bitradius$,$\bitlength$ and $\headradiusA$, set of milling directions $\millingDirSet$}
	\KwResult{Optimized version of $\partDomain$}
	\While{!converged}
	{
		EvaluateSensitivities($\partDomain$,Boundary conditions)\;
		$\speed$ = ComputeShapeDerivative()\;
		$\maxSpeed$ = max($\speed$)\;
		$\eta$ = MillingConstraint($\partDomain$,$\normal$,$\millingDirSet$,maxIters)\;
		\For{$\surfacePt$ on boundary of $\partDomain$}{
			\uIf{$\eta(\surfacePt) == 0$}{$\speed(\surfacePt) = \posVal\maxSpeed$}
			\Else{$\speed(\surfacePt) = \eta(\surfacePt)\speed(\surfacePt)$}
		}
		$\eps = $ LineSearch($\partDomain$,$\speed$)\;
		LevelSetAdvection($\partDomain$,$\eta\speed$,$\eps$)\;
	}
	LevelSetClose($\partDomain$,$\bitradius$)\;
	\caption{Inner Loop of the Augmented Lagrangian Algorithm for Level Set Topology Optimization --- CNC Constraint Relaxed Algorithm}
	\label{alg:relaxedAlgorithm}
\end{algorithm}
\normalsize

\bibliography{mybib}{}
\bibliographystyle{plain}

\end{document}